\documentclass[11pt]{article}

\usepackage[english]{babel}
\usepackage[utf8]{inputenc}
\usepackage{hyperref}
\usepackage{authblk}
\usepackage{graphicx}
\usepackage{mathpazo}
\usepackage{amsmath,amssymb}
\usepackage{setspace}
\usepackage{titlesec}
\usepackage{apacite}
\usepackage[letterpaper,margin=1in]{geometry}

\renewcommand\thesection{\arabic{section}.}
\renewcommand\thesubsection{\arabic{section}.\arabic{subsection}}

\titlespacing*{\section}
  {0pt}{0.5ex plus 1ex minus 0.2ex}{0.3ex plus 0.2ex}
\titlespacing*{\subsection}
  {0pt}{0.5ex plus 1ex minus 0.2ex}{0.3ex plus 0.2ex}
\titlespacing*{\subsubsection}
  {0pt}{0.5ex plus 1ex minus 0.2ex}{0.3ex plus 0.2ex}

\begin{document}

\title{A Geometry-Aware AI Emulator for the Coupled Whole Atmosphere from Earth’s Surface to the Ionosphere and Thermosphere}

\author[1]{Jiahui Hu}
\author[1,2]{Wenjun Dong\thanks{Corresponding author: \texttt{wenjun@gats-inc.com}}}

\affil[1]{Center for Space and Atmospheric Research, Embry-Riddle Aeronautical University, Daytona Beach, Florida, USA}
\affil[2]{Global Atmospheric Technologies and Sciences, Boulder, Colorado, USA}

\date{}
\maketitle

\section*{Abstract}
Whole-atmosphere models such as WACCM-X resolve coupling from the Earth surface to the Mesosphere-Lower-Thermosphere (MLT), and Ionosphere-Thermosphere (IT) systems with expensive computational costs. Here we introduce CAM-NET, a geometry-aware Spherical Fourier Neural Operator (SFNO) surrogate for emulating WACCM-X variability from Earth’s surface to IT region. CAM-NET is trained on 3-hourly WACCM-X simulations and predicts neutral winds, temperature, pressure-coordinate vertical velocity, electron density, and zonal ion drift. The framework combines a Spherical Fourier Neural Operator (SFNO) backbone with a newly developed lightweight module that extends the frozen atmospheric representation to plasma variables. For the held-out simulation, CAM-NET preserves the dominant IT morphology and remains stable during multi-day autoregressive rollouts. Spherical-harmonic diagnostics show that the model retains low-degree variability while damping high-wavenumber mesospheric structures, especially near 90 km where gravity wave breaks. CAM-NET is intended as a computationally efficient emulator of WACCM-X, rather than an operational forecasting system. These results demonstrate its potential for rapid ensemble experiments, uncertainty quantification, and sensitivity studies of large-scale coupled whole atmospheric variability.

\section{Introduction}
\label{sec:introduction}
Variability generated in the lower atmosphere can propagate upward through waves, tides, and large-scale circulation changes, influencing the thermosphere and ionosphere where space-weather impacts emerge. Capturing this vertical coupling requires models that span the atmosphere from the surface to near-Earth space. Physics-based systems such as the Whole Atmosphere Community Climate Model with Thermosphere-Ionosphere eXtension (WACCM-X) provide this capability by resolving coupled neutral-atmosphere and ionospheric processes, while operational weather systems such as the ECMWF Integrated Forecasting System (IFS) primarily target tropospheric and stratospheric dynamics. However, whole-atmosphere simulations remain computationally demanding, and these computational limitations directly affect practical space-weather applications. Uncertainty in thermospheric density, ionospheric electron density, and electrodynamic drifts propagates into satellite-drag estimation, radio-wave propagation, GNSS positioning, and studies of lower-atmosphere forcing on geospace variability. A fast emulator that reproduces the dominant coupled response of a whole-atmosphere model could enable large ensembles, uncertainty quantification, and sensitivity experiments that remain prohibitively expensive with full-physics simulations.

Data-driven weather models have recently demonstrated strong foreca. Examples include FourCastNet \cite{pathak2022fourcastnet}, GraphCast \cite{lam2023learning}, Pangu-Weather \cite{bi2023accurate}, FuXi \cite{chen2023fuxi, zhong2024fuxi}, FengWu \cite{chen2023fengwu}, ClimaX \cite{nguyen2023climax}, and GenCast \cite{price2023gencast}, with recent work also examining uncertainty quantification for data-driven forecasts \cite{bulte2025uncertainty}. These systems are largely trained on lower-atmosphere reanalysis data such as ERA5 \cite{hersbach2020era5}, the fifth-generation atmospheric reanalysis produced by ECMWF. Their success motivates the central question of this study: Can geometry-aware AI architectures emulate the coupled variability of the whole atmosphere? Extending AI weather models to this regime is nontrivial because the upper atmosphere is governed by changing composition, wave dissipation, electrodynamic forcing, and neutral-ionospheric coupling that are absent or secondary in conventional lower-atmosphere benchmarks. Beyond achieving low one-step prediction error, a successful surrogate must remain stable during autoregressive rollouts, reproduce physically meaningful variability across spatial scales, and transfer learned atmospheric representations to ionospheric variables.

The atmosphere exhibits distinct dynamical regimes across altitudes from surface to thermosphere, due to change in density, composition, and dominant forcing mechanisms \cite{liu2024assessment}. Mesoscale structures, with horizontal wavelengths of approximately 10–1000 km, such as gravity waves (GWs), originate primarily from tropospheric sources including deep convection and topography. These waves propagate vertically into the middle and upper atmosphere and eventually dissipate due to increasing molecular viscosity and background wind shear \cite{fritts2003gravity}. Wave dissipation generates secondary GWs with broader spatial scales and faster phase speeds, some of which can reach the Ionosphere-Thermosphere (IT) region, where they modulate neutral density and ionospheric plasma \cite{fritts2011gravity,vadas2009generation, dong2020SA,Fritts2020SA}. Observational evidence, such as electron perturbations, has been linked to these Mesospheric-Lower-Thermosphere (MLT) layers \cite{forbes2016gravity, azeem2015multisensor, liu2017medium}. Therefore, a unified framework spanning from the Earth surface to the IT region is essentially important for studying how lower-atmosphere variability projects into near-Earth space. The computational cost of full-physics models, however, motivates complementary AI surrogates that can reproduce large-scale coupled evolution rapidly while identifying where current architectures fail to retain small-scale processes. 

Here we introduce CAM-NET as a proof-of-concept AI framework for whole-atmosphere prediction across the neutral atmosphere and ionosphere, trained on WACCM-X dataset. The goal is not to replace WACCM-X, but to learn a fast surrogate of its coupled global evolution that can support rapid rollouts and future ensemble workflows. CAM-NET builds on the Spherical Fourier Neural Operator (SFNO) \cite{bonev2023spherical}, which represents global fields using spherical harmonics and therefore avoids geometric inconsistencies that can arise when planar Fourier methods are applied to latitude-longitude geophysical data. This geometry-aware design is important for prediction problems that span all longitudes, latitudes, and altitude regimes. Because the WACCM-X data is sampled at $0.9^\circ$ × $1.25^\circ$ and every fifth vertical level, the present study focuses on resolved large-scale and planetary-scale variability rather than fully resolved gravity-wave dynamics at the smallest mesoscale wavelengths. As with other finite spectral representations, however, the spherical-harmonic truncation also introduces a scale-selective smoothing that must be quantified.

A central design choice in CAM-NET is the newly developed lightweight module. Rather than retraining a full whole-atmosphere network for each new target variable, we keep the atmospheric backbone fixed and attach a lightweight ionospheric module for electron density and zonal ion drift. This module predicts temporal residual increments, reducing the dynamic range of the learning problem and improving stability during autoregressive rollouts. The resulting architecture provides a transferable pathway for extending an atmospheric AI backbone toward whole atmospheric couplings.

This study makes three testable contributions. First, it evaluates whether an SFNO surrogate can emulate WACCM-X neutral-atmosphere evolution, including wind components and temperature, across representative levels during independent autoregressive rollouts, then compared to the AFNO baseline. Second, it tests whether a frozen neutral-atmosphere backbone can support residual prediction of electron density and zonal ion drift. Third, it quantifies scale-dependent failure modes using spherical-harmonic spectra, showing where large-scale emulation remains reliable and where high-wavenumber mesospheric variability is damped.


\section{Method}
\label{sec:method}
The methods are designed to test whether a geometry-aware neural operator can serve as a stable surrogate for WACCM-X whole-atmosphere evolution. Sub-section \ref{subsect: WACCM-X data} describes the WACCM-X training and inferencing data, and subsection \ref{subsect:CAM-NET} describes the CAM-NET architecture, autoregressive training strategy, and ionospheric fine-tuning module.

\subsection{Data Source}
\label{subsect: WACCM-X data}
WACCM-X is a global atmospheric model that simulates dynamics from the Earth surface to the ionosphere \cite{liu2018development}. Nudged by the Modern-Era Retrospective Analysis for Research and Applications–version 2 (MERRA-v2), WACCM-X produces neutral atmospheric and ionospheric variables at a temporal resolution of 3~hours and a spatial resolution of $0.9^\circ \times 1.25^\circ$ (latitude $\times$ longitude), across different height levels defined by a hybrid sigma–pressure coordinate. This coordinate transitions from terrain-following sigma coordinates near the surface to pure pressure levels in the middle and upper atmosphere. To reduce dimensionality while preserving the large-scale vertical structure, every fifth height level is extracted from the native WACCM-X vertical grid and used for CAM-NET training and inference. Consequently, CAM-NET produces outputs at the same horizontal resolution as WACCM-X (roughly 100~km) and at a 3-hour temporal resolution. The resulting CAM-NET input tensor has dimensions (29 x 4) × 144 × 288 [(altitude x variable) x latitude x longitude]. The selected vertical levels correspond approximately to geometric altitudes, with diagnostic analyses shown at approximately 20, 90, and 250 km for neutral variables and 150, 250, and 350 km for ionospheric variables. 

The neutral prognostic targets are zonal wind U $[m/s]$, meridional wind $V [m/s]$, temperature $T [K]$, and pressure vertical velocity $\omega [Pa/s]$. Ionospheric targets are electron density $Ne [cm^{-3}]$ and zonal ion drift $UI [m/s]$. Static and external inputs include orography, sea-ice fraction, Kp, and F10.7 solar radio flux. We use daily mean F10.7 and 3-hourly Kp values consistent with the forcing used in WACCM-X. Because the present study focuses on the establishment of an AI surrogate of WACCM-X, all training and evaluation objectives are drawn from WACCM-X simulations. Accordingly, all skill metrics reported below measure emulation of WACCM-X, not observational forecast accuracy. Observational validation is outside the scope of this first surrogate-modeling study and is identified as a necessary next step.

Each 3-hourly NetCDF output file is stacked along the time dimension and aggregated into yearly HDF5 files for training and inference. Data from 2001–2011 are used for training, 2012–2013 for validation/testing, and 2014 for held-out inference. CAM-NET is initialized from a WACCM-X atmospheric state and then advanced autoregressively to predict subsequent 3-hourly states. This design evaluates whether the model can maintain stable coupled evolution when its own predictions are repeatedly recycled as inputs. The 2001–2011 training period contains 32,136 samples, the 2012–2013 validation/test period contains 5,848 samples, and the 2014 held-out inference period contains 2,920 3-hourly samples before excluding boundary samples required for input history and rollout construction. Normalization coefficients, climatologies, and any temporal smoothing are computed only from the training period. 

\subsection{CAM-NET Architecture and Training/Validation Strategy}
\label{subsect:CAM-NET}
CAM-NET leverages SFNO to model global atmospheric dynamics while reducing computational cost relative to full-physics WACCM-X simulations. Unlike conventional Fourier Neural Operators (FNOs), which transform fields in a Euclidean domain, SFNO replaces the discrete Fourier transform with the spherical harmonic transform (SHT) to preserve Earth’s spherical geometry \cite{bonev2023modelling,bonev2023spherical}. The SHT projects a spatial field $u(\theta, \phi)$ onto the spherical harmonic basis $Y_\ell^m$, yielding spectral coefficients:
\begin{equation}
    \hat{u}_{\ell m} = \mathcal{F}_s[u](\ell,m)
    = \int_{S^2} u(\theta,\phi) Y_{\ell}^{m*}(\theta,\phi)\,d\Omega,
\end{equation}

where $\ell$ and m denote spherical-harmonic degree and order. In spectral space, CAM-NET applies a learned complex-valued filter K through mode-wise multiplication,

\begin{equation}
    \hat{v}_{\ell m} = K_{\ell m}\hat{u}_{\ell m}. 
\end{equation}

The filtered signal is then transformed back to physical space using the inverse SHT,

\begin{equation}
    v(\theta,\phi) = \mathcal{F}_s^{-1}[\hat{v}](\theta,\phi)
    = \sum_{\ell=0}^{L}\sum_{m=-\ell}^{\ell}\hat{v}_{\ell m}Y_{\ell}^{m}(\theta,\phi).  
\end{equation}

Here $\theta$ is colatitude, $\phi$ is longitude, $K_{\ell m}$ is the learned spectral filter, and L is the maximum retained spherical-harmonic degree.

\begin{figure}[!t]
    \centering
   \includegraphics[width=\linewidth]{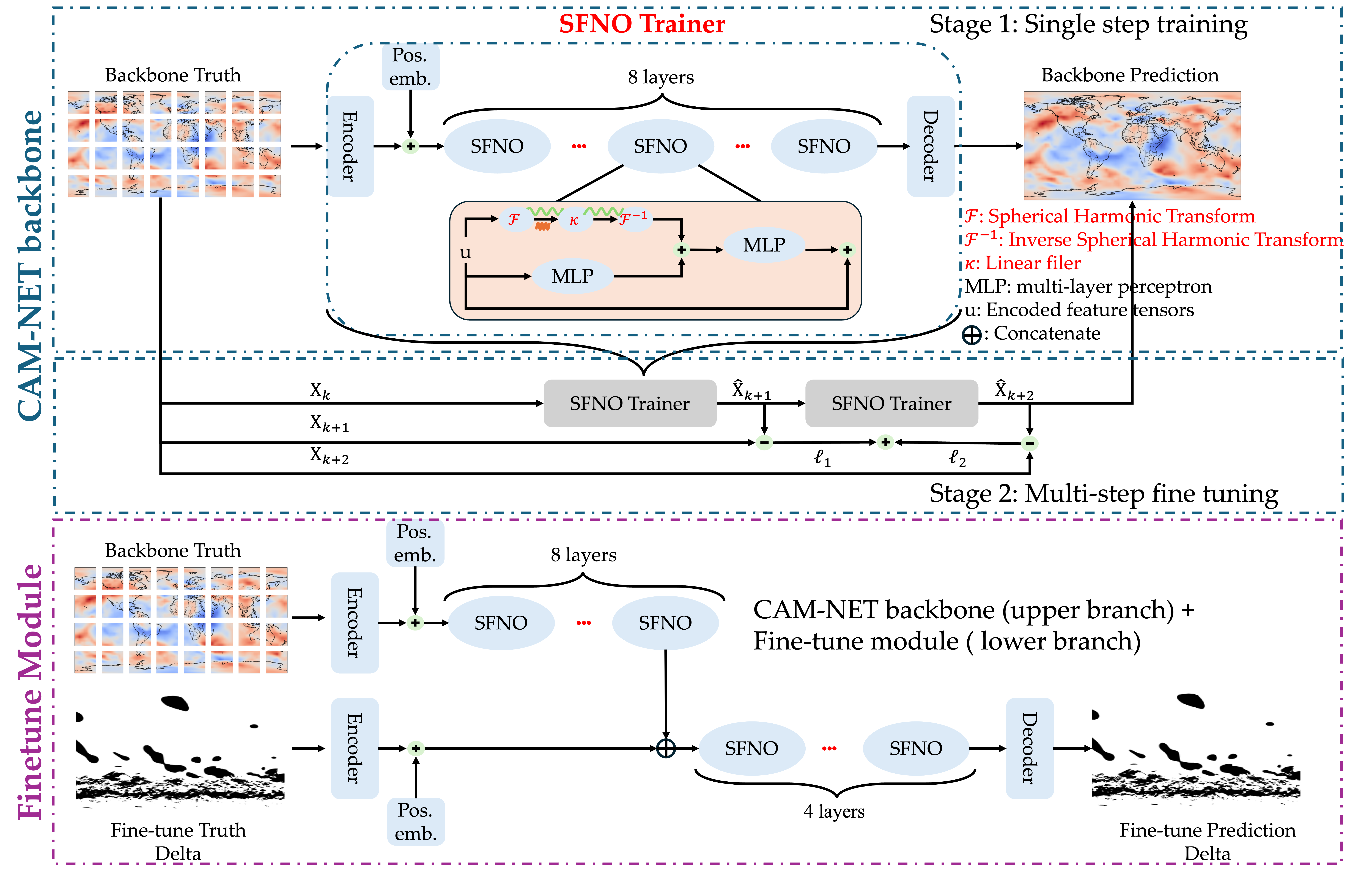}
    \caption{(top)Schematic representation of a Spherical Fourier Neural Operator (SFNO) backbone model for global geophysical data prediction. The model takes ground truth data, encodes it, and combines it with positional embeddings. The encoded feature tensors pass through multiple SFNO layers, which utilize a multi-head architecture with spherical harmonic transforms ($\mathcal{F}$) and inverse transforms ($\mathcal{F}^{-1}$) to process global-scale information efficiently. Each head applies a linear filter ($\kappa$) and a multi-layer perceptron (MLP) before aggregating the outputs. The decoder reconstructs the final global prediction, preserving spatial structures. (bottom) Ionospheric fine-tuning schema coupled with the CAM-NET backbone. Physical variables are projected into latent feature spaces and concatenated with ionospheric feature embeddings from a lower-capacity branch. The fused features are passed through additional SFNO blocks and decoded to predict temporal increments of ionospheric variables.}
    \label{fig:camnet}
\end{figure}
The CAM-NET backbone structure comprises three major components: (1) an input encoder that projects atmospheric variables into a latent space and incorporates positional embeddings to preserve spatial structure, (2) spectral convolutional layers based on SFNO that operate in the frequency domain, and (3) a decoder that reconstructs predictions in physical space. Each SFNO layer projects the input into spectral space, applies spectral filtering, and returns to physical space with residual connections and Multi-Layer Perceptrons (MLPs). 

To test whether CAM-NET can remain stable when used as a forecasting surrogate, the model is trained in two stages. In the first warm-start stage, it learns the single-step mapping from time t to t + $\delta t$. In the second stage, the model is rolled forward for K successive steps using its own predictions as inputs, and the loss is accumulated across the rollout. This closed-loop training reduces exposure bias and directly targets the error growth that arises during autoregressive whole-atmosphere prediction.

\begin{equation}
    \mathcal{L}_{multi-step} = \sum_{k=1}^{K} w_k ||\hat{\mathbf{u}}_{t+k} - \mathbf{u}_{t+k}||_2^2
\end{equation}

Figure \ref{fig:camnet} summarizes the architecture of the CAM-NET backbone. The process begins with input data (``truth’’), then passed through an encoder that integrates positional embeddings to preserve spatial structures. Each SFNO block centers around a Fourier transform pipeline $\mathcal{F}^{-1} \circ \kappa \circ \mathcal{F}$, where $\mathcal{F}$ denotes the SHT and $\kappa$ is a complex-valued, frequency-wise learnable filter. This block effectively performs global convolution on the sphere by manipulating spectral coefficients. To reduce memory usage and capture multiscale behavior, SFNO blocks also incorporate up- and down-scaling by truncating high-frequency modes in $\mathcal{F}$ or evaluating $\mathcal{F}^{-1}$ at higher spatial resolution. Point-wise MLPs follow each spectral operation to introduce local nonlinear feature mixing. The spherical-harmonic transform preserves the global spherical representation, although static fields and positional embeddings mean that the full network is not strictly rotation-equivariant. Finally, outputs from all SFNO layers are passed to a decoder that reconstructs the full-field prediction in physical space.

For whole-atmosphere AI models to be useful beyond a single diagnostic, they must be reusable across related prediction tasks without requiring full retraining. This requirement is especially important for thermosphere-ionosphere applications, where target variables may be sparse, highly variable, or tied to specific space-weather questions. CAM-NET therefore separates the learned atmospheric backbone from application-specific ionospheric modules, allowing the same large-scale atmospheric representation to condition multiple downstream prediction tasks.

Conventional whole-atmosphere models often couple a dynamical core to specialized ionospheric or diagnostic modules. CAM-NET adopts an analogous modular strategy in an AI framework. The pretrained atmospheric branch encodes the neutral state using the frozen SFNO backbone, while a lower-capacity ionospheric branch encodes electron-density and ion-drift information. These feature representations are concatenated and passed through four additional SFNO blocks to predict residual increments of the ionospheric variables. Because the atmospheric backbone remains frozen, ionospheric fine-tuning does not alter the neutral-atmosphere prediction and instead acts as a downstream module conditioned on the learned whole-atmosphere state. 

CAM-NET comprises a physical backbone and a lower-capacity ionospheric fine-tuning module, both with embedding dimension 384. The backbone uses 232 channels with $l_{max}/m_{max}$ =64/49 and approximately $8.2\times10^7$ trainable parameters, while the fine-tuning module uses 29 channels with $l_{max}/m_{max}$ =24/19 and approximately $1.8\times10^7$ parameters, reflecting its conditional prediction role. Each sample provides 9 hours of atmospheric history, and the model iteratively predicts the next 3-hour state. Training used Adam with an initial learning rate of $2.5\times10^{-4}$, StepLR scheduling, a global batch size of 8, z-score normalization from the physical training set, no explicit weight decay, and an absolute squared geometric $L_2$ loss with equal variable/channel and rollout-step weights. Checkpoints were selected by minimum validation loss on 2012–2013 rollouts. After warm start, training and validation losses steadily decreased and plateaued before inference, indicating convergence and supporting CAM-NET’s use as a low-cost surrogate for multi-day whole-atmosphere prediction.

To isolate the role of spherical geometry, we also trained an Adaptive Fourier Neural Operator (AFNO) baseline using the same WACCM-X training, validation, and held-out inference periods as CAM-NET. AFNO applies Fourier-domain token mixing on the latitude-longitude grid and therefore provides a planar spectral baseline against which the spherical-harmonic representation can be evaluated. The AFNO baseline used the same input variables, 9-h input history, 3-h prediction interval, normalization procedure, and autoregressive evaluation protocol as the SFNO backbone. This comparison is intended to test whether the geometry-aware spherical representation improves global whole-atmosphere emulation relative to a conventional Fourier operator applied directly to gridded geophysical fields. The AFNO baseline used embedding dimension 384, 8 layers, a global batch size of 8, and approximately $8.5\times10^7$ trainable parameters. The AFNO baseline was configured to match the SFNO backbone as closely as possible in input history, output variables, training data, and rollout protocol, so that the comparison primarily reflects the difference between planar Fourier and spherical-harmonic representations.

\section{Results}
\label{sec:results}
We evaluate CAM-NET by asking whether a single AI framework can preserve the dominant global structures of a coupled whole-atmosphere simulation across altitude regimes while remaining stable during autoregressive prediction. The evaluation has three parts: (1) Quantitative error analysis between WACCM-X and CAM-NET, and spectral diagnostics of scale-dependent skill, (2) direct benchmarking against a planar AFNO baseline, and (3) assessment of the residual ionospheric extension. Neutral variables are evaluated at approximately 20, 90, and 250 km, representing the lower stratosphere, mesopause region, and thermosphere, while ionospheric variables are evaluated at approximately 150, 250, and 350 km.

\subsection{Benchmark Against a Planar Fourier Baseline and 120-h Rollout Stability}
The AFNO comparison provides a direct test of whether the spherical representation contributes to CAM-NET skill. Because the AFNO baseline became numerically unstable during autoregressive integration beyond approximately 30 h, meaningful comparisons between the two architectures were limited to this forecast horizon. We therefore use the 30-h lead time as a common benchmark to isolate the impact of the spherical representation before evaluating the long-term autoregressive stability of CAM-NET separately. Table \ref{tbl:benchmark} shows that SFNO outperforms AFNO across all representative neutral variables and altitudes at the 30-h lead time. Relative to AFNO, SFNO reduces RMSE for every listed variable-altitude pair, with the largest gains appearing in lower-atmosphere horizontal winds and temperature, where AFNO produces weak or negative anomaly correlation while SFNO retains strong spatial pattern skill. Averaged across the neutral-variable benchmark, ACC increases from approximately 0.04 for AFNO to approximately 0.69 for SFNO. These results support the central design choice of CAM-NET: global whole-atmosphere emulation benefits from a spectral representation that respects spherical geometry. AFNO can efficiently model long-range dependencies, but its planar Fourier representation is less well matched to latitude-longitude fields spanning the poles, equator, and all longitudes. SFNO reduces this geometric mismatch by operating in a spherical-harmonic basis.

The improvement is not uniform, however. Pressure-coordinate vertical velocity remains less skillful than temperature and horizontal winds, especially at higher altitude, indicating that geometry-aware representation improves the baseline but does not fully solve the problem of intermittent wave-driven variability. Since the AFNO baseline could not maintain stable autoregressive integration beyond approximately 30 h, the long-term evaluation focuses exclusively on CAM-NET. We therefore assess whether the selected SFNO architecture remains stable during 120-h (5-day) autoregressive rollouts.

To assess rollout stability, we evaluate CAM-NET over 120-h autoregressive predictions on the held-out period. Bias, MAE, RMSE, and error standard deviation are averaged over the full rollout window, while the 120-h ACC summarizes pattern skill at the end of the 5-day integration, as Table \ref{tbl:120h-rollout}. CAM-NET retains high 120-h pattern skill for large-scale temperature and horizontal winds at 20 and 250 km, but skill decreases near 90 km and for pressure-coordinate vertical velocity. At 250 km, mean 120-h ACC is approximately 0.70 for zonal wind, 0.84 for meridional wind, and 0.78 for temperature, while pressure vertical velocity has lower mean ACC of approximately 0.28. At 90 km, where gravity-wave-related structures are more intermittent, mean 120-h ACC decreases to approximately 0.42 for zonal wind, 0.44 for meridional wind, and 0.53 for temperature. These results support the visual interpretation that CAM-NET preserves large-scale morphology but has reduced fidelity for mesospheric high-wavenumber variability.

The ionospheric variables show stronger 120-h pattern skill over the available rollouts. Mean 120-h ACC for electron density is approximately 0.96, 0.88, and 0.80 at 150, 250, and 350 km, respectively. Mean 120-h ACC for zonal ion drift is approximately 0.84–0.85 across the same altitude range. These values indicate that the lightweight residual ionospheric module preserves the dominant large-scale ionospheric morphology during 5-day rollouts, although localized extrema and sharp gradients are smoothed in the map comparisons.

Taken together, the benchmark and rollout results show two complementary strengths of CAM-NET. First, the geometry-aware SFNO backbone substantially improves neutral-atmosphere emulation relative to a planar AFNO baseline. Second, the same SFNO-based architecture remains stable during 120-h autoregressive integration, preserving coherent large-scale atmospheric and ionospheric structures even as fine-scale mesospheric variability becomes increasingly damped.

\subsection{Neutral Whole-Atmosphere Prediction Across Altitude Regimes}
\label{subsect: MLT & IT predictions}
\begin{figure}[htbp]
    \centering
    \includegraphics[width=\linewidth]{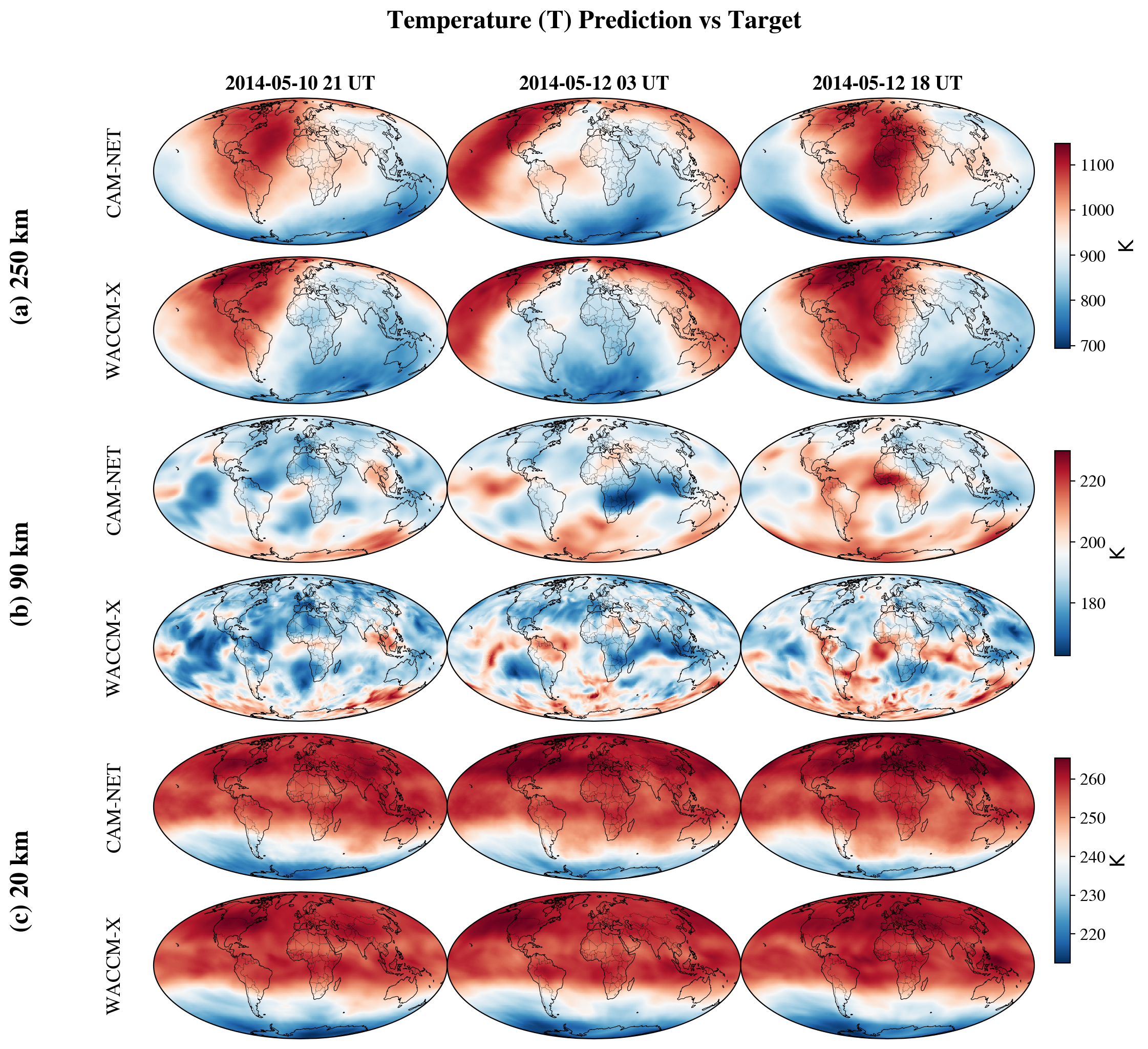}
    \caption{Temperature ($T [K]$) at (a) 250 km, (b) 90 km, and (c) 20 km at 21 UT on 10 May , 03 UT and 18 UT on 12 May 2014. In each panel, rows display results from CAM-NET (top) and WACCM-X (bottom).}
    \label{fig:2014-05-10-T}
\end{figure}

\begin{figure}[htbp]
    \centering
    \includegraphics[width=\linewidth]{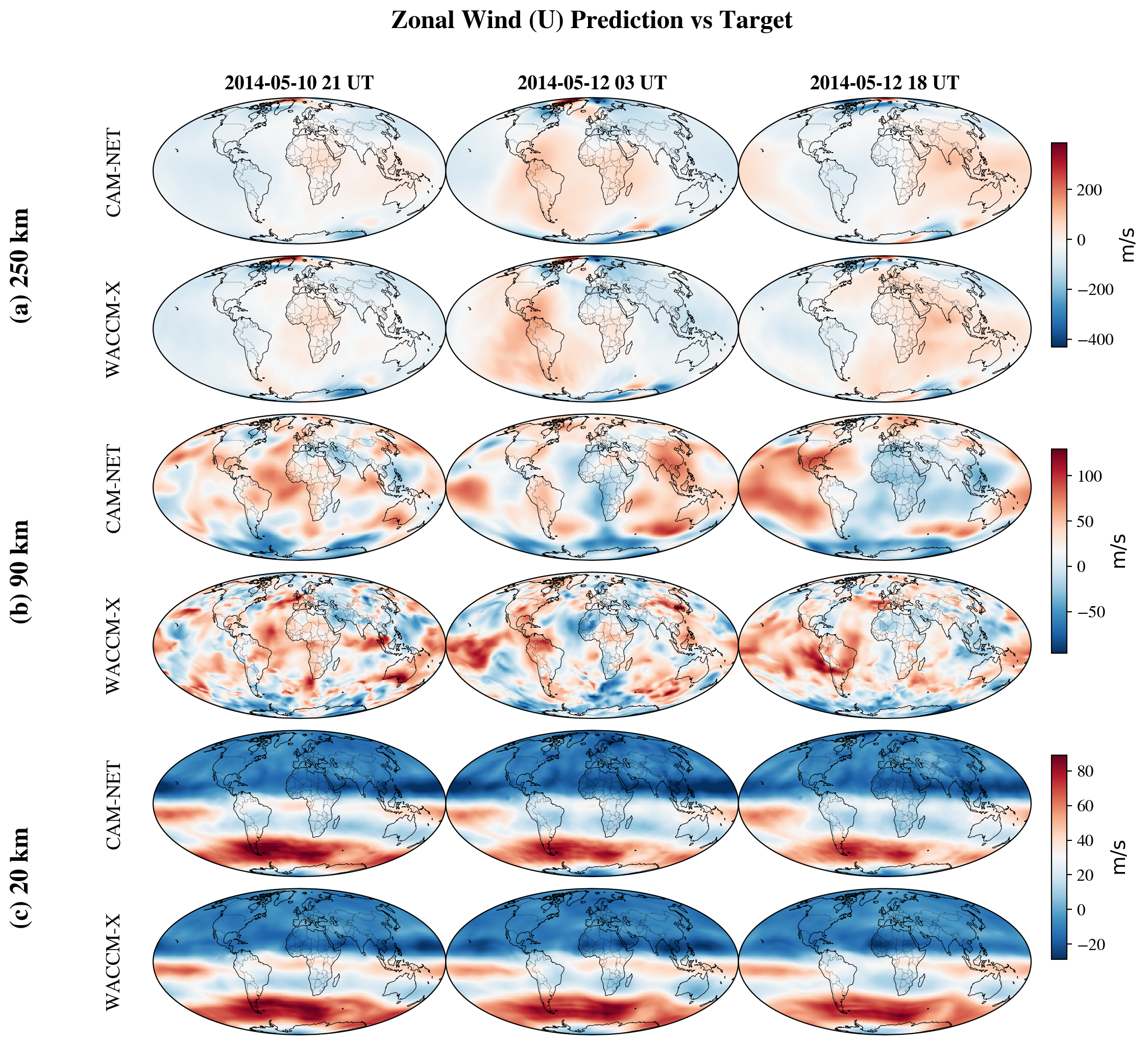}
    \caption{Zonal winds ($U[m/s]$) at (a) 250 km, (b) 90 km, and (c) 20 km at 21 UT on May 10th, 03 UT and 18 UT on May 12th 2014. In each panel, rows display results from CAM-NET (top) and WACCM-X (bottom).}
    \label{fig:2014-05-10-U}
\end{figure}

\begin{figure}[htbp]
    \centering
    \includegraphics[width=\linewidth]{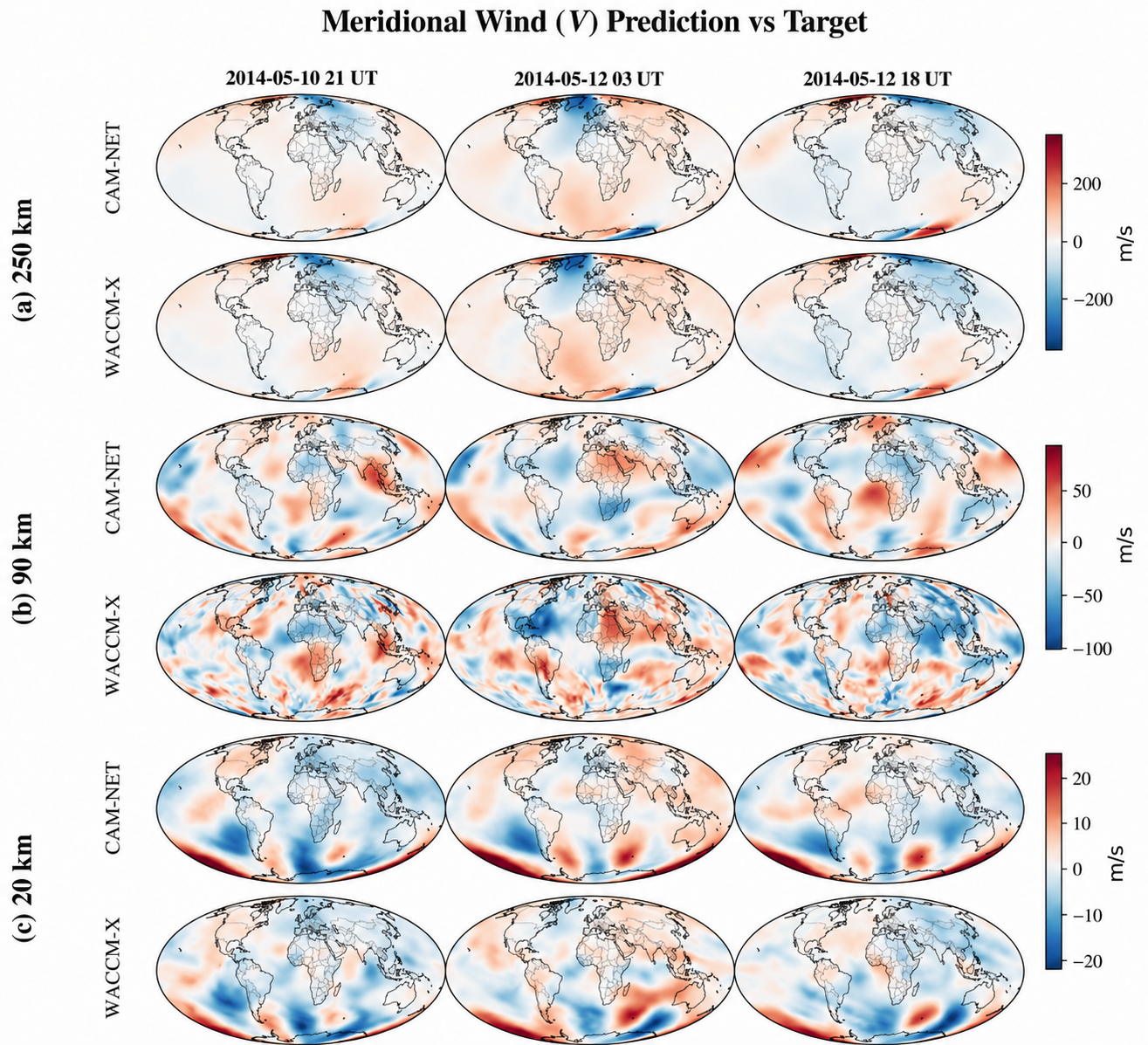}
    \caption{Meridional winds ($V[m/s]$) at (a) 250 km, (b) 90 km, and (c) 20 km at 21 UT on May 10th, 03 UT and 18 UT on May 12th 2014. In each panel, rows display results from CAM-NET (top) and WACCM-X (bottom).}
    \label{fig:2014-05-10-V}
\end{figure}

\begin{figure}[htbp]
    \centering
    \includegraphics[width=\linewidth]{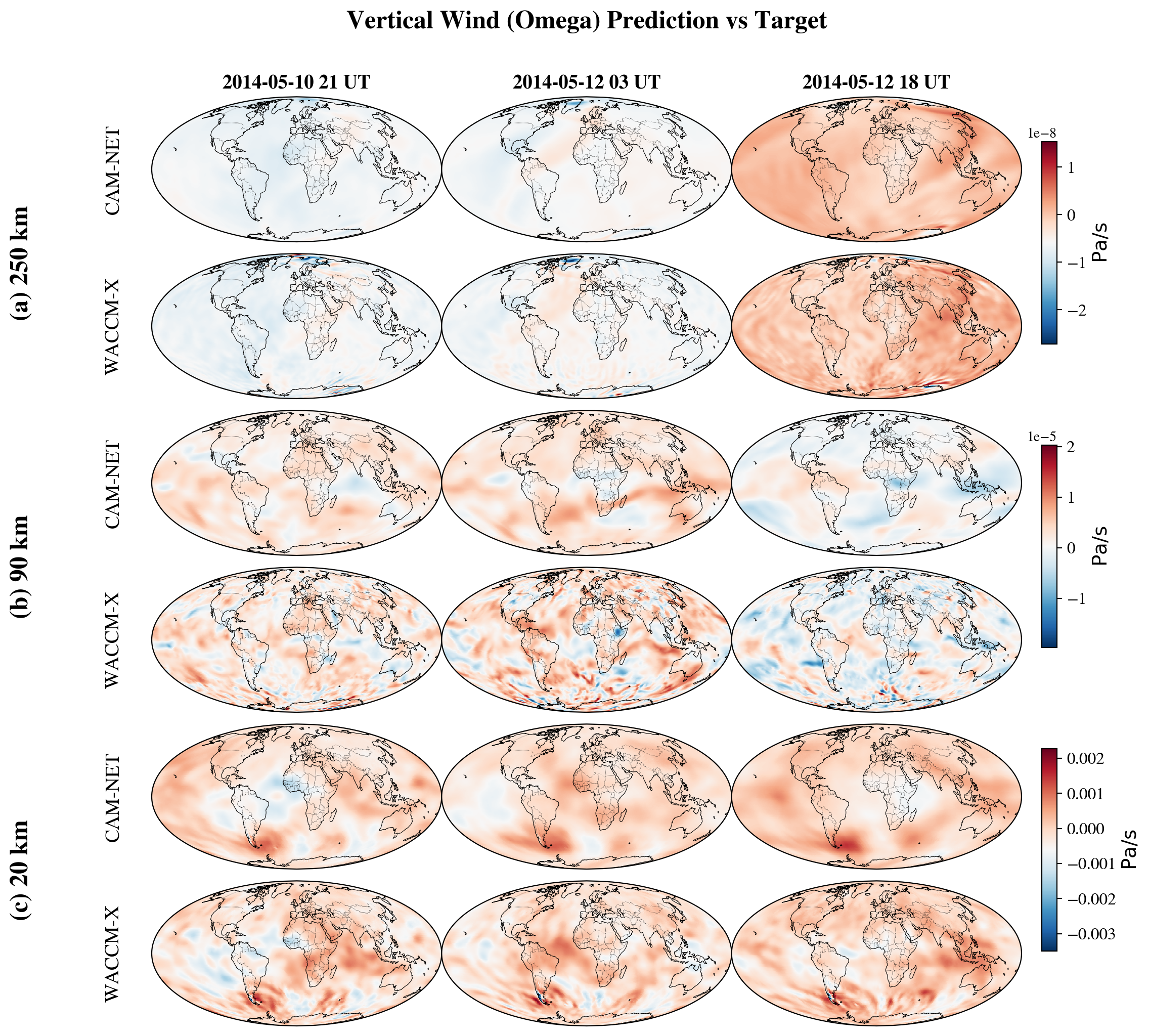}
    \caption{Pressure-coordinate vertical velocity ($Omega[pa/s]$) at (a) 250 km, (b) 90 km, and (c) 20 km at 21 UT on May 10th, 03 UT and 18 UT on May 12th 2014. In each panel, rows display results from CAM-NET (top) and WACCM-X (bottom).}
    \label{fig:2014-05-10-W}
\end{figure}

\begin{figure}[htbp]
    \centering
    \includegraphics[width=\linewidth]{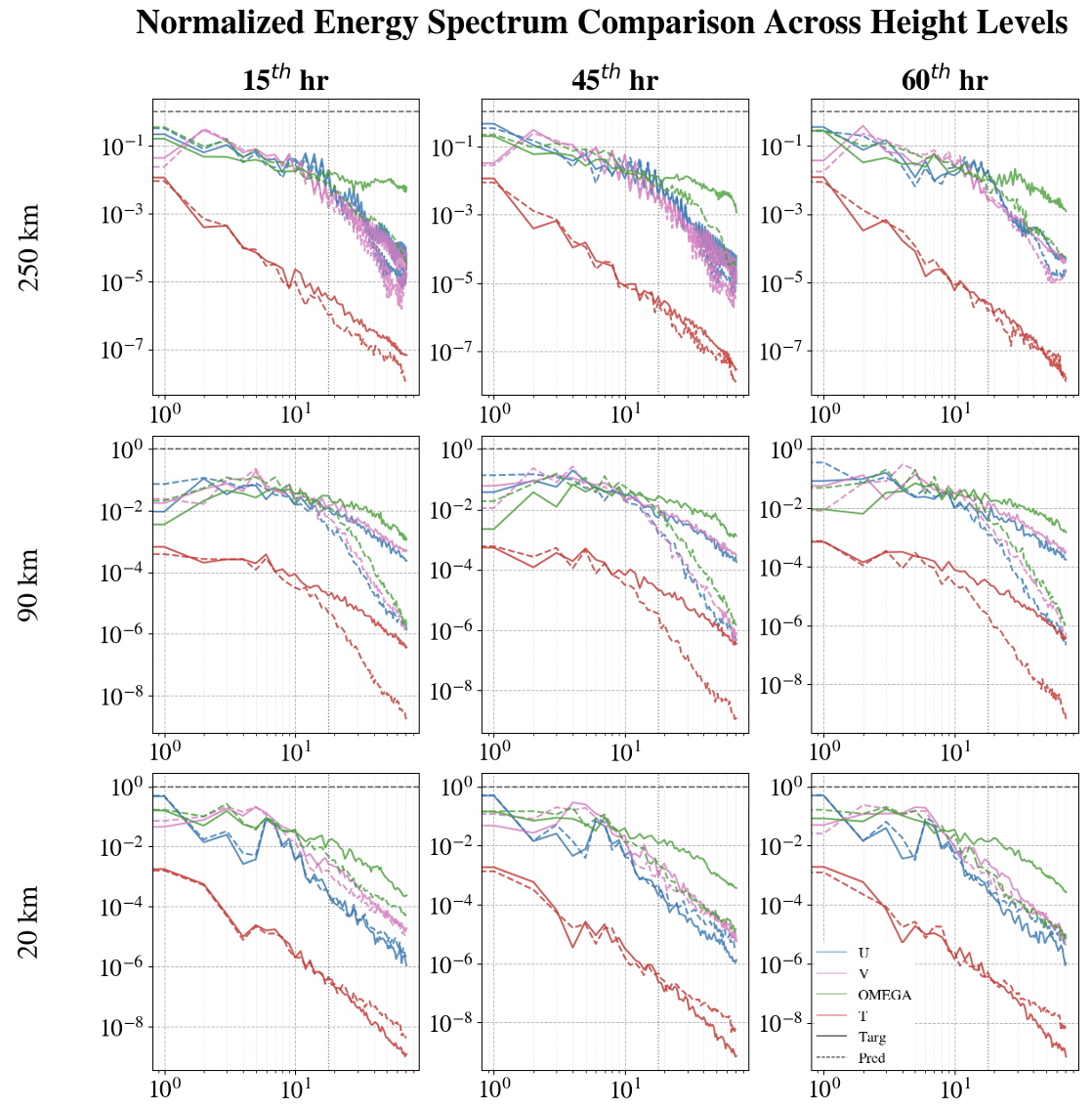}
    \caption{Spectra are averaged globally and normalized by total spectral energy at each altitude. Energy spectrum comparison at three different altitudes: 250 km (top), 90 km (middle) and 20 km (bottom), showing the normalized spectrum energy as a function of spherical harmonic degrees ($l$). The solid lines represent target values, while the dashed lines plot the predicted values for different variables: zonal wind ($U$) in blue, meridional wind ($V$) in magenta, vertical velocity ($\omega$) in green, and temperature ($T$). A vertical dashed line marks a reference point for large and small scale separation.}
    \label{fig:spectrum_comparison}
\end{figure}

The snapshots in Figures \ref{fig:2014-05-10-T} – \ref{fig:UI} were selected from a representative quiet-time interval. To avoid over-interpreting individual examples, all visual comparisons are accompanied by domain-wide metrics computed over the held-out evaluation period. Figure \ref{fig:2014-05-10-T} compares CAM-NET temperature predictions with WACCM-X target fields at 250, 90, and 20 km during the held-out 2014 rollout. CAM-NET captures the dominant large-scale temperature structures at all three altitudes, including smooth lower-stratospheric gradients and the day-night thermospheric contrast. The largest discrepancy occurs near 90 km, where WACCM-X contains patchy and filamentary structures associated with gravity-wave activity and wave breaking. CAM-NET smooths much of this variability. This behavior identifies an important boundary of the present approach: the model is effective for planetary-scale and synoptic-scale whole-atmosphere structure, but it does not yet fully retain localized mesospheric variability generated by intermittent wave processes.

Figures \ref{fig:2014-05-10-U}–\ref{fig:2014-05-10-W} extend this comparison to zonal wind, meridional wind, and pressure-coordinate vertical velocity. At 20 km, CAM-NET reproduces coherent large-scale circulation patterns and latitudinal gradients. At 250 km, it captures the dominant thermospheric wind morphology, including broad flow reversals and planetary-scale structures, although spatial shifts and amplitude differences increase. The largest visual mismatch again occurs near 90 km, where CAM-NET preserves the background circulation but smooths the patchy and filamentary structures present in WACCM-X. This limitation is most pronounced for vertical motion, which is particularly sensitive to small-scale wave dynamics and nonlinear momentum deposition. Because $\omega$ is a pressure-coordinate vertical velocity, its magnitude and interpretation vary strongly with altitude and background pressure. We therefore evaluate $\omega$ using both absolute errors and normalized errors relative to WACCM-X variability at each level.

Spherical-harmonic spectra are computed from normalized anomalies after removing the global mean at each level and time. The spectra are area-weighted and normalized by the total spectral energy at each altitude. Figure \ref{fig:spectrum_comparison} shows representative spectra at 15-, 45-, and 60-h lead times, while the summary statistics are computed over the held-out evaluation samples. We define low-degree variability as $\ell \leq 20$, corresponding approximately to horizontal scales larger than 2000 km, and high-degree variability as $\ell > 20$. CAM-NET reproduces the large-scale portion of the spectrum at low spherical-harmonic degrees across altitudes and lead times. For $\ell \leq 20$, the mean relative spectral error remains below 4.3\% at 20 km and 4.8\% at 250 km. At 90 km, agreement is maintained at the largest scales, but mesoscale energy is increasingly underestimated at higher degrees, especially for vertical velocity and temperature. Thus, CAM-NET is stable and accurate for planetary-scale whole-atmosphere structure, but its fidelity decreases for localized high-wavenumber variability.

The under-representation of small-scale energy is not only a model error but also a diagnostic of how current neural-operator architectures represent multiscale geophysical variability. CAM-NET retains large-scale, globally coherent structures but damps localized high-wavenumber features, consistent with both neural-network spectral bias \cite{rahaman2019spectral,khodakarami2025mitigating} and the finite truncation of spherical-harmonic representations. This result suggests that future AI whole-atmosphere models will likely require hybrid strategies, such as multiresolution operators, stochastic parameterizations, or physics-informed correction modules, to recover intermittent mesospheric processes while retaining the efficiency of global spectral architectures.

\subsection{Modular Extension to Ionospheric Electron Density and Zonal Ion Drift}

\begin{figure}
    \centering
    \includegraphics[width=\linewidth]{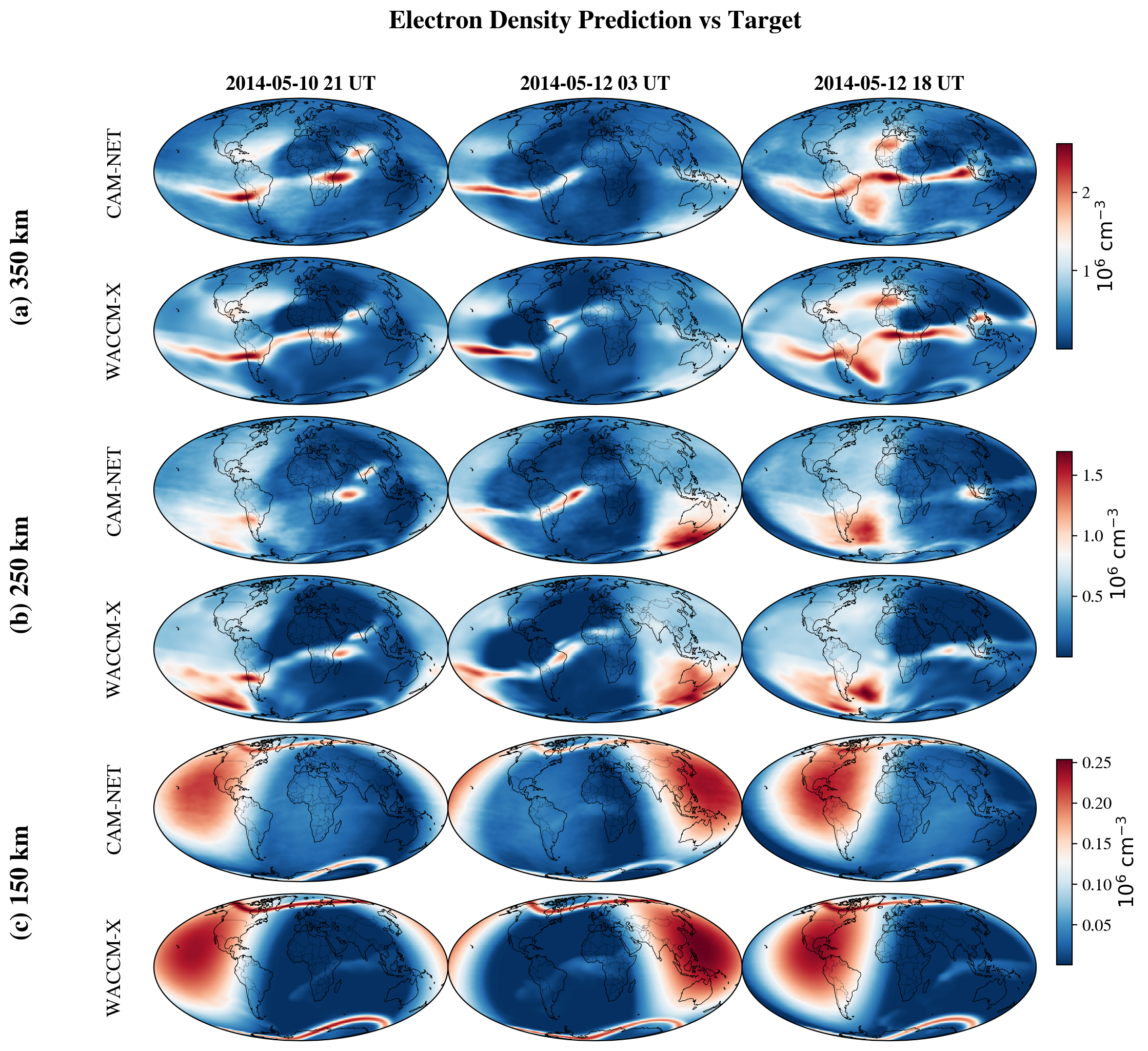}
    \caption{Global distribution of Electron Density [$10^6 cm^{-3}$] at (a) 350 $km$, (b) 250 $km$, (c) 150 $km$ for three selected times (2014-05-10 21 UT, 2014-05-12 03 UT, and 2014-05-12 18 UT).}
    \label{fig:Edens}
\end{figure}

Figure \ref{fig:Edens} evaluates whether the frozen atmospheric backbone can support ionospheric prediction through the residual fine-tuning module. CAM-NET reproduces the dominant electron-density morphology at 350, 250, and 150 km, including day-night asymmetry, equatorial plasma enhancements, and longitudinal variability associated with IT coupling. At 350 km, the model preserves the broad equatorial electron-density structure but underestimates localized maxima and sharp gradients. At 250 km, it captures the evolving large-scale plasma distribution, although mesoscale structures become increasingly diffuse with lead time. At 150 km, where spatial gradients are smoother, CAM-NET shows stronger agreement with the WACCM-X background ionization structure. These results show that the atmospheric backbone contains information useful for ionospheric prediction when coupled to a lightweight residual module.

\begin{figure}[htbp]
    \centering
    \includegraphics[width=\linewidth]{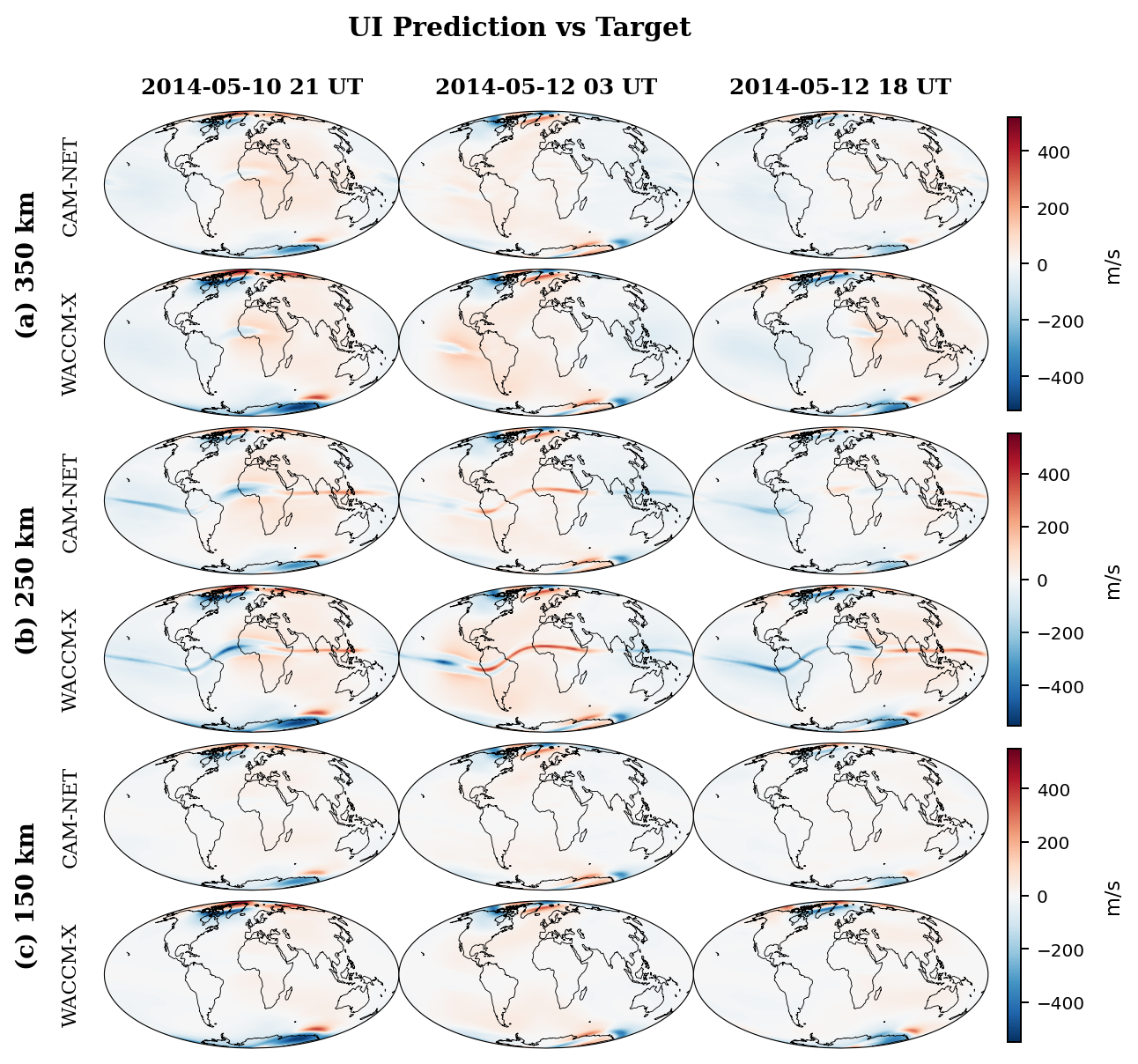}
    \caption{Global distribution of zonal ion drifts $[m/s]$ at (a) 350 $km$, (b) 250 $km$, (c) 150 $km$ for three selected times (2014-05-10 21 UT, 2014-05-12 03 UT, and 2014-05-12 18 UT).}
    \label{fig:UI}
\end{figure}

Figure \ref{fig:UI} compares CAM-NET zonal ion drift predictions with WACCM-X reference fields at 350, 250, and 150 km. Across these altitudes, CAM-NET reproduces the dominant large-scale electrodynamic morphology, including longitudinally extended drift bands, high-latitude variability, and hemispheric asymmetries. The coherent spatial organization across autoregressive rollouts indicates that residual prediction stabilizes the ionospheric extension even when the model repeatedly uses its own prior predictions as input.

At 350 km, where ion drift is strongly influenced by coupled IT electrodynamics and geomagnetic forcing, CAM-NET captures the primary drift polarity and broad spatial organization but smooths localized extrema, particularly in auroral and subauroral regions. Similar behavior occurs at 250 km, where broad equatorial and midlatitude structures are retained but embedded smaller-scale variability is attenuated. At 150 km, predicted drift structures are smoother and lower in amplitude, consistent with both reduced ion-drift magnitudes and the tendency of spectral neural operators to preserve large-scale variability preferentially. The gradual loss of small-scale structure mirrors the neutral-atmosphere spectral damping identified above. Importantly, however, the large-scale ion-drift morphology remains coherent across rollout times, showing that residual fine-tuning can transfer information from the neutral atmospheric backbone into ionospheric prediction while preserving the computational advantages of a frozen whole-atmosphere model. 

\begin{table}[htbp]
\centering
\caption{Time-averaged error metrics and 120-h lead-time pattern skill for neutral and ionospheric variables over the held-out autoregressive rollouts. Bias, MAE, RMSE, and error standard deviation are computed over all evaluated longitudes, latitudes, and times at each representative altitude; Bias is CAM-NET minus WACCM-X. The 120-h ACC summarizes pattern skill from available saved 5-day autoregressive rollout curves. Parentheses indicate the range of 120-h ACC across available rollouts. Electron-density errors are reported in units of $10^{6}$ cm$^{-3}$.}
\label{tab:combined_metrics}
\small
\resizebox{\textwidth}{!}{%
\begin{tabular}{llrrrrc}
\hline
Variable & Altitude [km] & Bias & MAE & RMSE & Error STD & 120-h ACC \\
\hline
$\omega$ [Pa s$^{-1}$] & 250 & $3.05{\times}10^{-11}$ & $1.76{\times}10^{-9}$ & $2.85{\times}10^{-9}$ & $2.84{\times}10^{-9}$ & 0.28 (0.04--0.37) \\
$\omega$ [Pa s$^{-1}$] & 90  & $9.87{\times}10^{-8}$  & $1.99{\times}10^{-6}$ & $2.66{\times}10^{-6}$ & $2.66{\times}10^{-6}$ & 0.36 (0.32--0.46) \\
$\omega$ [Pa s$^{-1}$] & 20  & $1.26{\times}10^{-5}$  & $2.04{\times}10^{-4}$ & $2.70{\times}10^{-4}$ & $2.70{\times}10^{-4}$ & 0.37 (0.32--0.41) \\

$T$ [K] & 250 & -8.13 & 49.17 & 56.60 & 55.38 & 0.78 (0.55--0.95) \\
$T$ [K] & 90  &  4.06 &  6.54 &  8.07 &  6.95 & 0.53 (0.48--0.60) \\
$T$ [K] & 20  &  2.10 &  2.86 &  3.78 &  3.12 & 0.74 (0.55--0.92) \\

$U$ [m s$^{-1}$] & 250 & -0.30 & 32.81 & 44.82 & 44.16 & 0.70 (0.55--0.80) \\
$U$ [m s$^{-1}$] & 90  & -1.36 & 18.37 & 23.22 & 23.14 & 0.42 (0.29--0.62) \\
$U$ [m s$^{-1}$] & 20  & -3.56 &  5.44 &  7.07 &  6.08 & 0.85 (0.74--0.93) \\

$V$ [m s$^{-1}$] & 250 & -2.78 & 24.63 & 40.69 & 40.51 & 0.84 (0.71--0.89) \\
$V$ [m s$^{-1}$] & 90  & -1.58 & 18.47 & 23.31 & 23.09 & 0.44 (0.33--0.54) \\
$V$ [m s$^{-1}$] & 20  &  0.86 &  3.42 &  4.52 &  4.41 & 0.58 (0.49--0.76) \\

$N_e$ [$10^{6}$ cm$^{-3}$] & 350 & -0.058 & 0.195 & 0.271 & 0.262 & 0.80 (0.76--0.90) \\
$N_e$ [$10^{6}$ cm$^{-3}$] & 250 & -0.016 & 0.108 & 0.148 & 0.146 & 0.88 (0.85--0.90) \\
$N_e$ [$10^{6}$ cm$^{-3}$] & 150 & -0.003 & 0.020 & 0.025 & 0.025 & 0.96 (0.96--0.97) \\

$U_i$ [m s$^{-1}$] & 350 & 8.58 & 48.22 & 70.05 & 69.29 & 0.85 (0.81--0.94) \\
$U_i$ [m s$^{-1}$] & 250 & 7.41 & 39.50 & 66.12 & 65.46 & 0.85 (0.80--0.93) \\
$U_i$ [m s$^{-1}$] & 150 & 5.23 & 32.84 & 58.93 & 53.77 & 0.84 (0.80--0.93) \\
\hline
\end{tabular}%
}
\label{tbl:120h-rollout}
\end{table}

\begin{table*}[htbp]
\centering
\caption{Neutral-atmosphere benchmark comparing the planar AFNO baseline with CAM-NET/SFNO. Bias, MAE, RMSE, and Error STD are averaged over the evaluated autoregressive rollout period, while ACC is reported at a 30-h lead time. Bias is computed as model minus WACCM-X reference. Lower absolute Bias, MAE, RMSE, and Error STD indicate smaller errors, while higher ACC indicates better spatial anomaly-pattern agreement.}
\label{tab:afno_sfno_metrics}
\small
\resizebox{\textwidth}{!}{%
\begin{tabular}{llrrrrrrrrrr}
\hline
& & \multicolumn{5}{c}{\textbf{AFNO}} & \multicolumn{5}{c}{\textbf{SFNO}} \\
\cline{3-7}\cline{8-12}
Variable & Altitude [km] & Bias & MAE & RMSE & Error STD & ACC & Bias & MAE & RMSE & Error STD & ACC \\
\hline
$\omega$ [Pa s$^{-1}$] & 250 & $1.06\times10^{-9}$ & $3.30\times10^{-9}$ & $7.95\times10^{-9}$ & $7.88\times10^{-9}$ & 0.018 & $-3.51\times10^{-11}$ & $1.57\times10^{-9}$ & $3.25\times10^{-9}$ & $3.25\times10^{-9}$ & 0.204 \\
$\omega$ [Pa s$^{-1}$] & 90  & $-2.21\times10^{-7}$ & $3.93\times10^{-6}$ & $8.88\times10^{-6}$ & $8.87\times10^{-6}$ & 0.046 & $6.97\times10^{-8}$ & $1.89\times10^{-6}$ & $2.53\times10^{-6}$ & $2.53\times10^{-6}$ & 0.484 \\
$\omega$ [Pa s$^{-1}$] & 20  & $-9.05\times10^{-4}$ & $1.05\times10^{-3}$ & $2.76\times10^{-3}$ & $2.61\times10^{-3}$ & 0.028 & $3.67\times10^{-6}$ & $2.45\times10^{-4}$ & $3.33\times10^{-4}$ & $3.33\times10^{-4}$ & 0.506 \\
\hline
$T$ [K] & 250 & -84.78 & 100.36 & 235.28 & 219.48 & -0.371 & 6.03 & 48.66 & 58.34 & 58.03 & 0.919 \\
$T$ [K] & 90  & 17.20 & 20.00 & 45.06 & 41.65 & 0.172 & 2.45 & 5.65 & 7.55 & 7.14 & 0.633 \\
$T$ [K] & 20  & -5.35 & 11.94 & 27.99 & 27.47 & 0.150 & 0.76 & 2.02 & 2.72 & 2.61 & 0.921 \\
\hline
$U$ [m s$^{-1}$] & 250 & 2.70 & 39.73 & 71.79 & 71.74 & 0.207 & 8.18 & 30.62 & 44.08 & 43.32 & 0.699 \\
$U$ [m s$^{-1}$] & 90  & 7.19 & 63.76 & 122.28 & 122.07 & 0.033 & -0.62 & 15.67 & 20.54 & 20.53 & 0.631 \\
$U$ [m s$^{-1}$] & 20  & 19.06 & 83.37 & 250.18 & 249.45 & -0.037 & -1.14 & 4.32 & 6.14 & 6.03 & 0.957 \\
\hline
$V$ [m s$^{-1}$] & 250 & -59.14 & 72.98 & 174.25 & 163.91 & 0.063 & -1.50 & 17.96 & 34.72 & 34.69 & 0.765 \\
$V$ [m s$^{-1}$] & 90  & 43.25 & 55.45 & 101.99 & 92.37 & 0.103 & -0.27 & 16.59 & 21.65 & 21.65 & 0.640 \\
$V$ [m s$^{-1}$] & 20  & -30.74 & 31.37 & 68.72 & 61.46 & 0.047 & -0.26 & 3.27 & 5.03 & 5.02 & 0.862 \\
\hline
\end{tabular}%
}
\label{tbl:benchmark}
\end{table*}




\section{Discussion and Summary}
\label{sec:discussion-summary}
CAM-NET demonstrates that a geometry-aware neural operator can emulate the large-scale response of WACCM-X across both neutral and ionospheric variables during multi-day autoregressive rollouts. Its primary value is as a computationally efficient surrogate rather than as a replacement for WACCM-X. Once trained, CAM-NET can rapidly generate WACCM-X-like realizations for ensemble design, sensitivity analysis, and hypothesis testing, allowing computationally intensive full-physics simulations to be focused on the most informative cases.

Direct comparisons with state-of-the-art AI weather models are inherently limited because those systems primarily target the lower atmosphere and do not extend into the IT system. CAM-NET is therefore complementary to models such as FourCastNet, GraphCast, and Pangu-Weather. Its contribution is not improved medium-range weather forecasting, but the demonstration that geometry-aware neural operators can emulate variability across the coupled whole atmosphere and ionosphere.

The AFNO baseline clarifies what aspect of CAM-NET is most important for this application. The improvement of SFNO over AFNO verifies that treating the latitude-longitude grid as a flat periodic domain is not optimal for vertically extended global geophysical fields, especially for multi-variables across whole atmospheric region. The spherical-harmonic representation better matches the geometry of WACCM-X output and improves both amplitude and spatial-pattern metrics for neutral variables. However, the comparison also shows that geometry awareness alone is insufficient for full multiscale fidelity. SFNO still damps high-wavenumber mesospheric structures, implying that future progress will require not only spherical geometry but also architectures or losses designed to retain intermittent small-scale variability.

A second contribution is modularity. By freezing the atmospheric backbone and training a lower-capacity residual ionospheric branch, CAM-NET separates the learned representation of whole-atmosphere dynamics from application-specific prediction targets. This design mirrors the modular philosophy of physics-based modeling systems, where dynamical cores are coupled to specialized parameterizations or diagnostic modules. In future applications, the same strategy could be tested for additional upper-atmosphere diagnostics, provided that there are suitable training targets and that the frozen backbone is shown to retain the relevant physical information.

Two limitations are particularly important. First, CAM-NET has a
reduced representation of small-scale mesospheric variability. Near
90~km, where gravity waves break, interact nonlinearly, and generate
secondary waves and turbulence, the model systematically
underestimates high-degree spectral energy and smooths localized
structures. This behavior likely reflects the finite truncation of
spherical harmonics, the tendency of $L_2$ losses to favor conditional
means, and the reduced vertical resolution used for training.
Consequently, the current model should be interpreted as a surrogate
for large-scale whole-atmosphere variability rather than as a complete
representation of mesospheric dynamics.

Second, the storm-time evaluation in Appendix~B reveals reduced
fidelity under strongly disturbed geomagnetic conditions. CAM-NET
remains numerically stable and retains portions of the broad
thermosphere--ionosphere organization, but it does not reproduce the
location, amplitude, and temporal evolution of localized storm-time
electron-density enhancements and ion-drift structures with the same
fidelity achieved during quieter conditions. This limitation likely
reflects the use of scalar Kp and F10.7 inputs, which do not fully
describe the spatial distribution and time history of magnetospheric
forcing, as well as the limited representation of extreme events in
the training record. The reduced state representation and frozen
atmospheric backbone may further limit the recovery of rapidly
evolving electrodynamic and composition responses.

The most direct next experiments are: (1) increase $l_{\max}$ and
$m_{\max}$ to quantify the resolution--skill tradeoff; (2) train with
spectral or gradient-weighted losses to penalize high-wavenumber
damping; (3) test multiresolution spherical operators or spherical
wavelets; (4) add stochastic residual models to represent unresolved
mesospheric variability; (5) evaluate whether finer vertical sampling
improves gravity-wave-related structures near 90~km; (6) increase the
representation and weighting of geomagnetically disturbed intervals
during training; and (7) test whether additional storm-relevant drivers or state variables
improve the representation of disturbed thermosphere--ionosphere conditions.

In summary, CAM-NET emulates large-scale WACCM-X
whole-atmosphere--ionosphere variability during multi-day
autoregressive rollouts. The AFNO benchmark shows that the spherical
neural-operator formulation substantially improves
neutral-atmosphere emulation relative to a planar Fourier baseline,
supporting the use of geometry-aware architectures in global geospace
applications. CAM-NET retains useful large-scale pattern skill for
temperature, horizontal winds, electron density, and zonal ion drift,
while pressure-coordinate vertical velocity, high-wavenumber
mesospheric structures, and strongly disturbed storm-time conditions
remain more challenging. In particular, the March 2015 storm
evaluation shows that the model can remain numerically stable while
exhibiting substantial errors in the location, amplitude, and timing
of localized ionospheric and electrodynamic structures. The current
system should therefore be interpreted as a fast emulator of
large-scale WACCM-X behavior, primarily under quiet to moderately
disturbed conditions, rather than as an observationally validated
operational space-weather forecast model. These results provide a
foundation for future ensemble, sensitivity, and
uncertainty-propagation studies, while also identifying extreme-event
generalization as a central target for further development.

\section*{Acknowledgments}
This work was supported by NSF Grant AGS-2327914 and NASA Grant 80NSSC24K0124. All training and inference tasks were performed on the NSF NCAR Derecho supercomputer. We gratefully acknowledge the provision of GPU resources, which were instrumental in enabling this research.

\section*{Author Contributions}

Jiahu Hu and Wenjun Dong contributed to the conceptualization, formal analysis, investigation, software development, visualization, original draft preparation, and review and editing of the manuscript. Wenjun Dong was additionally responsible for supervision and funding acquisition.

\section*{Data and Code Availability}
The WACCM-X training datasets are available through the NCAR Research Data Archive \cite{cisl_rda_dsd651034}. Six animations accompany this study and are available at \url{https://doi.org/10.5281/zenodo.18393643}. The code package is uploaded on Github: \url{https://github.com/Multi-Scale-Wave-Dynamics-Group/CAMNET-code.git}

\section*{Competing Interests}
The authors declare that they have no known competing financial interests or personal relationships that could have influenced the work reported in this paper.

\bibliographystyle{apacite}
\bibliography{CAMNET}

@article{azeem2015multisensor,
  title={Multisensor profiling of a concentric gravity wave event propagating from the troposphere to the ionosphere},
  author={Azeem, Irfan and Yue, Jia and Hoffmann, Lars and Miller, Steven D and Straka III, William C and Crowley, Geoff},
  journal={Geophysical research letters},
  volume={42},
  number={19},
  pages={7874--7880},
  year={2015},
  publisher={Wiley Online Library}
}

@inproceedings{bonev2023spherical,
  title={Spherical fourier neural operators: Learning stable dynamics on the sphere},
  author={Bonev, Boris and Kurth, Thorsten and Hundt, Christian and Pathak, Jaideep and Baust, Maximilian and Kashinath, Karthik and Anandkumar, Anima},
  booktitle={International conference on machine learning},
  pages={2806--2823},
  year={2023},
  organization={PMLR}
}

@inproceedings{bonev2023modelling,
  title={Modelling atmospheric dynamics with spherical fourier neural operators},
  author={Bonev, Boris and Kurth, Thorsten and Hundt, Christian and Pathak, Jaideep and Baust, Maximilian and Kashinath, Karthik and Anandkumar, Anima},
  booktitle={ICLR 2023 Workshop on Tackling Climate Change with Machine Learning, https://www. climatechange. ai/papers/iclr2023/47 (last access: 24 August 2023)},
  year={2023}
}

@article{bi2023accurate,
  title={Accurate medium-range global weather forecasting with 3D neural networks},
  author={Bi, Kaifeng and Xie, Lingxi and Zhang, Hengheng and Chen, Xin and Gu, Xiaotao and Tian, Qi},
  journal={Nature},
  volume={619},
  number={7970},
  pages={533--538},
  year={2023},
  publisher={Nature Publishing Group}
}

@article{bulte2025uncertainty,
  title={Uncertainty quantification for data-driven weather models},
  author={B{\"u}lte, Christopher and Horat, Nina and Quinting, Julian and Lerch, Sebastian},
  journal={Artificial Intelligence for the Earth Systems},
  year={2025},
  publisher={American Meteorological Society}
}

@misc{cisl_rda_dsd651034,
 author = "F. {Gasperini}",
 title = "SD WACCM-X v2.1",
 publisher  = "Research Data Archive at the National Center for Atmospheric Research, Computational and Information Systems Laboratory",
 address = {Boulder CO},
 year  = "2025",
 url = "https://rda.ucar.edu/datasets/d651034/"
}

@article{chen2023fengwu,
  title={Fengwu: Pushing the skillful global medium-range weather forecast beyond 10 days lead},
  author={Chen, Kang and Han, Tao and Gong, Junchao and Bai, Lei and Ling, Fenghua and Luo, Jing-Jia and Chen, Xi and Ma, Leiming and Zhang, Tianning and Su, Rui and others},
  journal={arXiv preprint arXiv:2304.02948},
  year={2023}
}

@article{chen2023fuxi,
  title={FuXi: A cascade machine learning forecasting system for 15-day global weather forecast},
  author={Chen, Lei and Zhong, Xiaohui and Zhang, Feng and Cheng, Yuan and Xu, Yinghui and Qi, Yuan and Li, Hao},
  journal={npj climate and atmospheric science},
  volume={6},
  number={1},
  pages={190},
  year={2023},
  publisher={Nature Publishing Group UK London}
}

@article{dong2020SA,
    author = {Dong, Wenjun and Fritts, David C. and Lund, Thomas S. and Wieland, Scott A. and Zhang, Shaodong},
    title = {Self-Acceleration and Instability of Gravity Wave Packets: 2. Two-Dimensional Packet Propagation, Instability Dynamics, and Transient Flow Responses},
    journal = {Journal of Geophysical Research: Atmospheres},
    volume = {125},
    number = {3},
    pages = {e2019JD030691},
    doi = {https://doi.org/10.1029/2019JD030691},
    year = {2020}
}

@article{fritts2003gravity,
  title={Gravity wave dynamics and effects in the middle atmosphere},
  author={Fritts, David C and Alexander, M Joan},
  journal={Reviews of geophysics},
  volume={41},
  number={1},
  year={2003},
  publisher={Wiley Online Library}
}

@incollection{fritts2011gravity,
  title={Gravity wave influences in the thermosphere and ionosphere: Observations and recent modeling},
  author={Fritts, David C and Lund, Thomas S},
  booktitle={Aeronomy of the Earth's Atmosphere and Ionosphere},
  pages={109--130},
  year={2011},
  publisher={Springer}
}

@article{Fritts2020SA,
    author = {Fritts, David C. and Dong, Wenjun and Lund, Thomas S. and Wieland, Scott and Laughman, Brian},
    title = {Self-Acceleration and Instability of Gravity Wave Packets: 3. Three-Dimensional Packet Propagation, Secondary Gravity Waves, Momentum Transport, and Transient Mean Forcing in Tidal Winds},
    journal = {Journal of Geophysical Research: Atmospheres},
    volume = {125},
    number = {3},
    pages = {e2019JD030692},
    doi = {https://doi.org/10.1029/2019JD030692},
    year = {2020}
}

@article{forbes2016gravity,
  title={Gravity wave-induced variability of the middle thermosphere},
  author={Forbes, Jeffrey M and Bruinsma, Sean L and Doornbos, Eelco and Zhang, Xiaoli},
  journal={Journal of Geophysical Research: Space Physics},
  volume={121},
  number={7},
  pages={6914--6923},
  year={2016},
  publisher={Wiley Online Library}
}

@article{hersbach2020era5,
  title={The ERA5 global reanalysis, quarterly journal of the royal meteorological society},
  author={Hersbach, H and Bell, B and Berrisford, P and Hirahara, S and Hor{\'a}nyi, A and Mu{\~n}oz-Sabater, J and Nicolas, J and Peubey, C and Radu, R and Schepers, D},
  year={2020},
  publisher={John Wiley \& Sons Ltd on behalf of the Royal Meteorological Society}
}

@article{khodakarami2025mitigating,
  title={Mitigating Spectral Bias in Neural Operators via High-Frequency Scaling for Physical Systems},
  author={Khodakarami, Siavash and Oommen, Vivek and Bora, Aniruddha and Karniadakis, George Em},
  journal={arXiv preprint arXiv:2503.13695},
  year={2025}
}

@article{liu2017medium,
  title={Medium-scale gravity wave activity in the bottomside F region in tropical regions},
  author={Liu, Huixin and Pedatella, Nicholas and Hocke, Klemens},
  journal={Geophysical Research Letters},
  volume={44},
  number={14},
  pages={7099--7105},
  year={2017},
  publisher={Wiley Online Library}
}

@article{liu2018development,
  title={Development and validation of the whole atmosphere community climate model with thermosphere and ionosphere extension (WACCM-X 2.0)},
  author={Liu, Han-Li and Bardeen, Charles G and Foster, Benjamin T and Lauritzen, Peter and Liu, Jing and Lu, Gang and Marsh, Daniel R and Maute, Astrid and McInerney, Joseph M and Pedatella, Nicholas M and others},
  journal={Journal of Advances in Modeling Earth Systems},
  volume={10},
  number={2},
  pages={381--402},
  year={2018},
  publisher={Wiley Online Library}
}

@article{lam2023learning,
  title={Learning skillful medium-range global weather forecasting},
  author={Lam, Remi and Sanchez-Gonzalez, Alvaro and Willson, Matthew and Wirnsberger, Peter and Fortunato, Meire and Alet, Ferran and Ravuri, Suman and Ewalds, Timo and Eaton-Rosen, Zach and Hu, Weihua and others},
  journal={Science},
  volume={382},
  number={6677},
  pages={1416--1421},
  year={2023},
  publisher={American Association for the Advancement of Science}
}

@article{liu2024assessment,
  title={Assessment of gravity waves from tropopause to thermosphere and ionosphere in high-resolution WACCM-X simulations},
  author={Liu, H-L and Lauritzen, PH and Vitt, F and Goldhaber, S},
  journal={Journal of Advances in Modeling Earth Systems},
  volume={16},
  number={6},
  pages={e2023MS004024},
  year={2024},
  publisher={Wiley Online Library}
}

@article{nguyen2023climax,
  title={Climax: A foundation model for weather and climate},
  author={Nguyen, Tung and Brandstetter, Johannes and Kapoor, Ashish and Gupta, Jayesh K and Grover, Aditya},
  journal={arXiv preprint arXiv:2301.10343},
  year={2023}
}

@article{price2023gencast,
  title={Gencast: Diffusion-based ensemble forecasting for medium-range weather},
  author={Price, Ilan and Sanchez-Gonzalez, Alvaro and Alet, Ferran and Andersson, Tom R and El-Kadi, Andrew and Masters, Dominic and Ewalds, Timo and Stott, Jacklynn and Mohamed, Shakir and Battaglia, Peter and others},
  journal={arXiv preprint arXiv:2312.15796},
  year={2023}
}

@article{pathak2022fourcastnet,
  title={Fourcastnet: A global data-driven high-resolution weather model using adaptive fourier neural operators},
  author={Pathak, Jaideep and Subramanian, Shashank and Harrington, Peter and Raja, Sanjeev and Chattopadhyay, Ashesh and Mardani, Morteza and Kurth, Thorsten and Hall, David and Li, Zongyi and Azizzadenesheli, Kamyar and others},
  journal={arXiv preprint arXiv:2202.11214},
  year={2022}
}

@inproceedings{rahaman2019spectral,
  title={On the spectral bias of neural networks},
  author={Rahaman, Nasim and Baratin, Aristide and Arpit, Devansh and Draxler, Felix and Lin, Min and Hamprecht, Fred and Bengio, Yoshua and Courville, Aaron},
  booktitle={International conference on machine learning},
  pages={5301--5310},
  year={2019},
  organization={PMLR}
}

@article{vadas2009generation,
  title={Generation of large-scale gravity waves and neutral winds in the thermosphere from the dissipation of convectively generated gravity waves},
  author={Vadas, Sharon L and Liu, Han-li},
  journal={Journal of Geophysical Research: Space Physics},
  volume={114},
  number={A10},
  year={2009},
  publisher={Wiley Online Library}
}

@article{zhong2024fuxi,
  title={FuXi-2.0: Advancing machine learning weather forecasting model for practical applications},
  author={Zhong, Xiaohui and Chen, Lei and Fan, Xu and Qian, Wenxu and Liu, Jun and Li, Hao},
  journal={arXiv preprint arXiv:2409.07188},
  year={2024}
}

\appendix
\renewcommand{\thesubsection}{\thesection\arabic{subsection}}

\renewcommand{\thefigure}{S\arabic{figure}}

\setcounter{figure}{0}

\section{Additional AFNO--SFNO baseline diagnostics}

\begin{figure}[h]
    \centering
    \includegraphics[width=\linewidth]{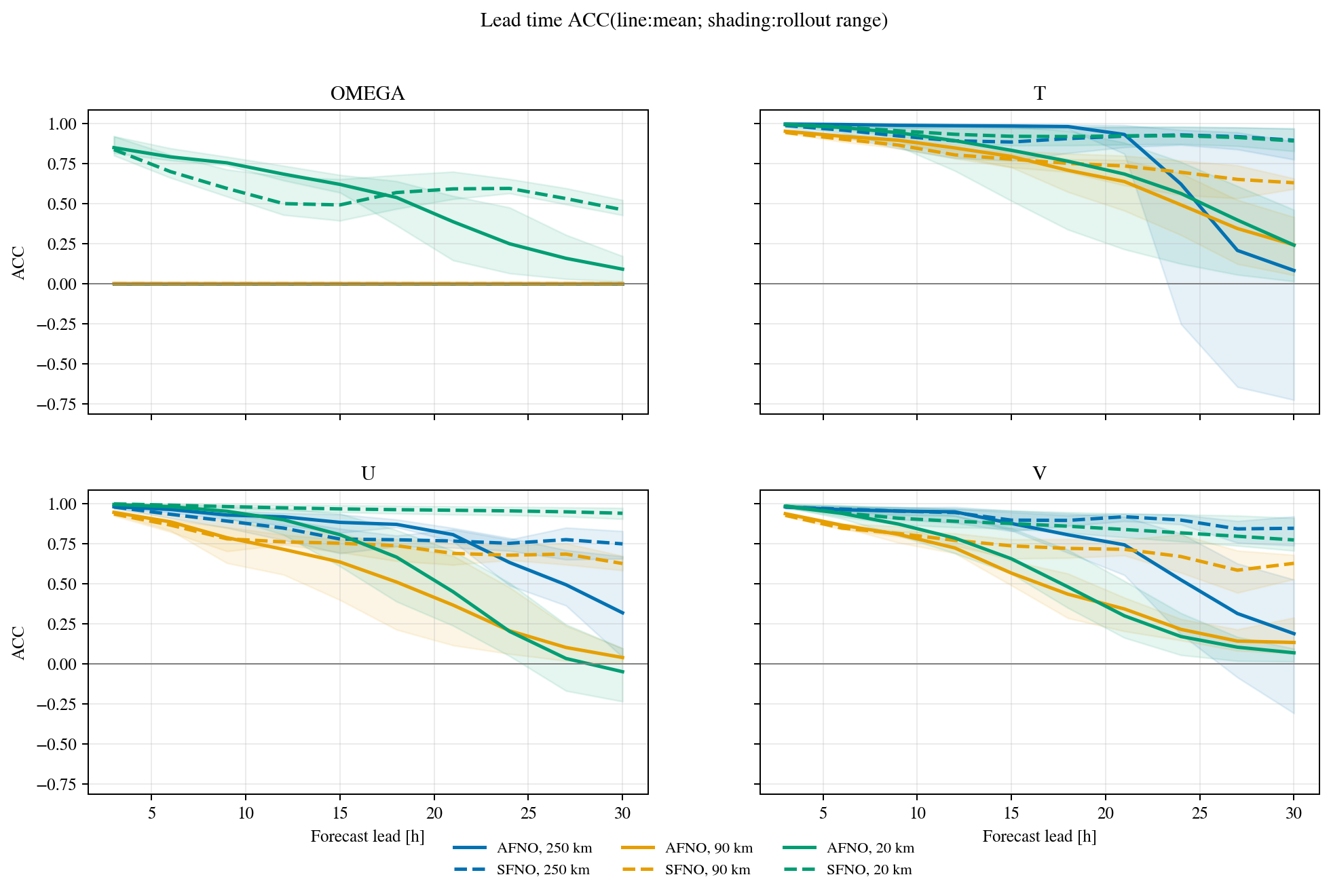}
    \caption{Lead-time dependence of the anomaly correlation coefficient (ACC) for AFNO and SFNO neutral-atmosphere rollouts. ACC is shown for pressure-coordinate vertical velocity $\omega$, temperature $T$, zonal wind $U$, and meridional wind $V$ at approximately 250, 90, and 20~km. Solid lines denote AFNO, dashed lines denote SFNO/CAM-NET, and colors indicate altitude. Lines show the mean ACC across the evaluated autoregressive rollouts, while shading shows the rollout range. The horizontal gray line marks zero ACC. Higher ACC indicates better spatial anomaly-pattern agreement with WACCM-X. SFNO generally maintains higher and more stable ACC than AFNO, especially for temperature and horizontal winds, while $\omega$ remains the most challenging variable.}
    \label{fig:baseline-lead-time}
\end{figure}

Figure~\ref{fig:baseline-lead-time} extends the AFNO--SFNO comparison from the summary metrics in the main text to the full lead-time evolution of the anomaly correlation coefficient (ACC). The comparison is organized by neutral variable--pressure-coordinate vertical velocity $\omega$, temperature $T$, zonal wind $U$, and meridional wind $V$--and by representative altitude. For temperature and horizontal winds, SFNO generally maintains higher and more stable ACC than AFNO as lead time increases, especially at 20 and 90~km. AFNO can retain useful short-lead skill but undergoes rapid autoregressive error growth and becomes unstable after approximately 30~h. The comparison is therefore limited to the common 30-h interval. SFNO preserves coherent large-scale structure over this interval and remains stable in the 120-h rollouts evaluated in the main text. Pressure-coordinate vertical velocity remains the most difficult variable, particularly at 250 and 90~km, consistent with its sensitivity to intermittent wave-driven and small-scale motion.

\begin{figure}[h]
    \centering
    \includegraphics[width=0.8\linewidth]{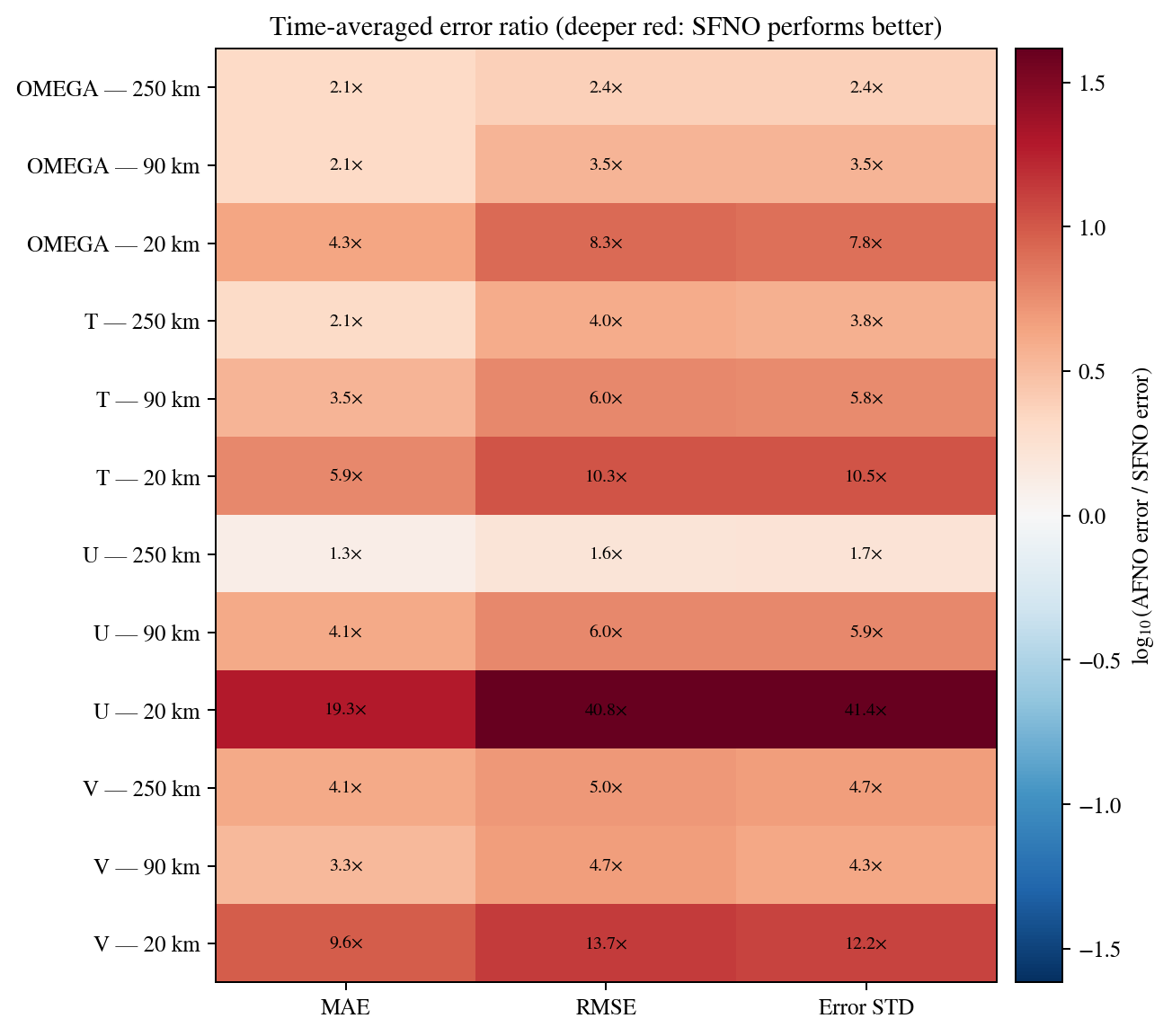}
    \caption{Time-averaged AFNO-to-SFNO error ratios for neutral-atmosphere variables. Rows show pressure-coordinate vertical velocity $\omega$, temperature $T$, zonal wind $U$, and meridional wind $V$ at approximately 250, 90, and 20~km; columns show MAE, RMSE, and the standard deviation of the model--reference error. Cell annotations give the linear ratio, defined as AFNO error divided by SFNO error. The color scale shows $\log_{10}(\mathrm{AFNO\ error}/\mathrm{SFNO\ error})$. Ratios greater than one indicate smaller SFNO errors. All plotted ratios exceed one, showing that SFNO reduces the time-averaged amplitude error for every variable, altitude, and error metric shown.}
    \label{fig:baseline-error-ratio}
\end{figure}

Figure~\ref{fig:baseline-error-ratio} summarizes the same comparison using time-averaged amplitude-error ratios. Each cell reports the AFNO error divided by the corresponding SFNO error for mean absolute error (MAE), root-mean-square error (RMSE), and error standard deviation. Every ratio exceeds one, showing that SFNO reduces time-averaged error for every variable, altitude, and metric included in the comparison. The largest improvement occurs for lower-atmosphere horizontal winds, particularly $U$ at 20~km, where AFNO RMSE and error standard deviation exceed the SFNO values by more than a factor of 40. Temperature and meridional wind also show substantial improvements, with ratios commonly between approximately 4 and 14 depending on altitude and metric. The smallest relative improvement occurs for $U$ at 250~km, where SFNO nevertheless reduces the errors by factors of approximately 1.3--1.7. Together, Figs.~\ref{fig:baseline-lead-time} and~\ref{fig:baseline-error-ratio} show that the spherical-harmonic representation improves both pattern retention and amplitude accuracy relative to the planar AFNO baseline, although the benefit varies by variable and altitude.

\section{CAM-NET Performance and Limitations During a Geomagnetic Storm}
\label{app:storm}
Geomagnetic storms represent an especially challenging regime for whole-atmosphere emulation because the thermosphere–ionosphere system undergoes rapid, nonlinear, and spatially localized changes. We therefore evaluate CAM-NET during the March 2015 geomagnetic storm as an extreme-event stress test. Unlike the quiet-time evaluation in the main text, this analysis is intended primarily to identify the limitations of the current model rather than to demonstrate equivalent performance to WACCM-X. The storm-period rollout was initialized from a WACCM-X state at 09:00 UTC on 16 March 2015. The snapshots at 21:00 UTC on 16 March and at 03:00 and 18:00 UTC on 18 March correspond to lead times of 12, 42, and 57~h, respectively, and sample pre-storm and storm-recovery conditions.

\subsection{Neutral-atmosphere winds}
\label{app:storm-neutral}

\begin{figure}[h]
    \centering
    \includegraphics[width=\linewidth]{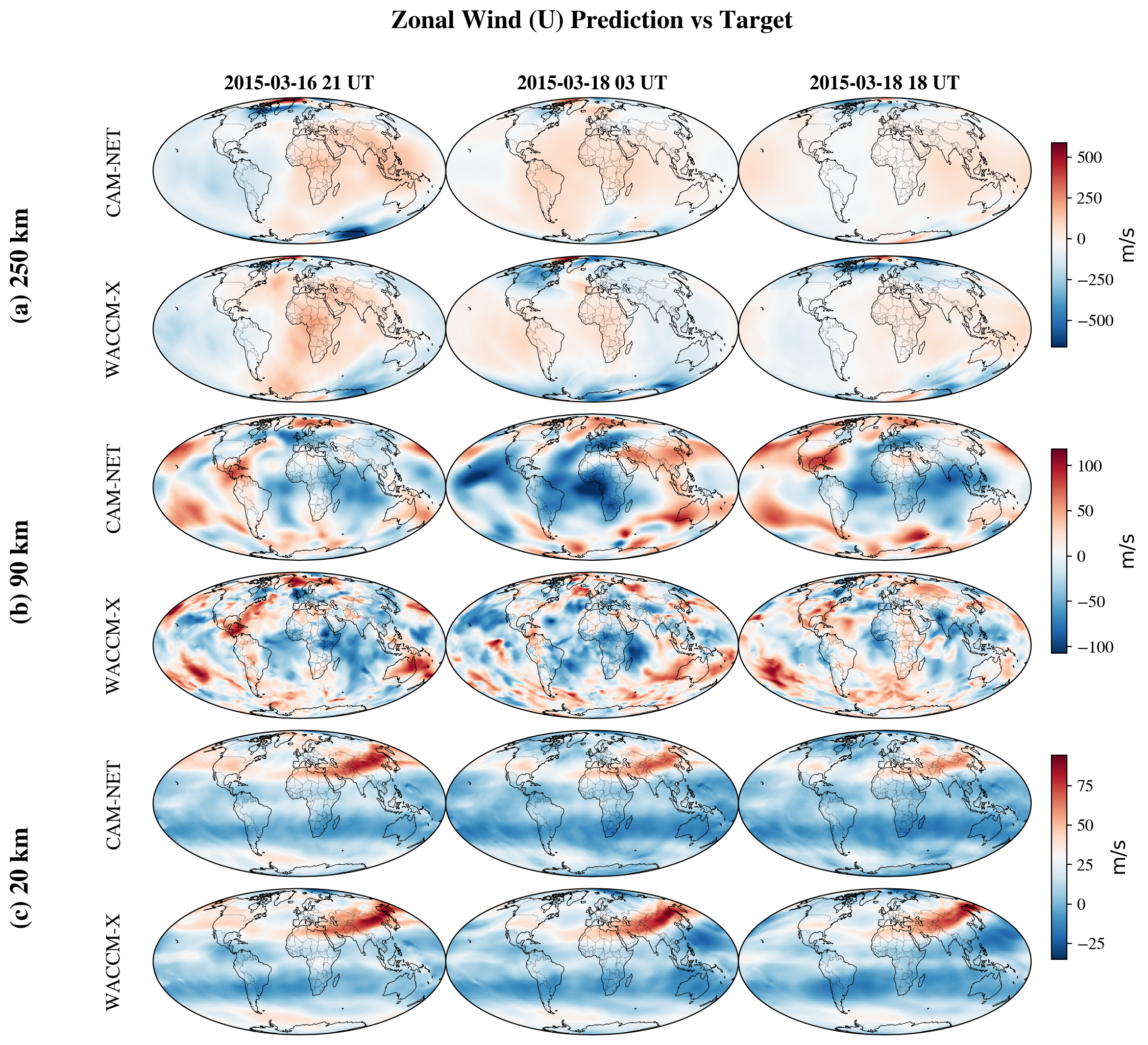}
    \caption{Zonal wind ($U$, m~s$^{-1}$) at approximately (a) 250~km, (b) 90~km, and (c) 20~km. Columns show 21:00~UT on 16 March, 03:00~UT on 18 March, and 18:00~UT on 18 March 2015. Within each panel, the upper row shows CAM-NET and the lower row shows WACCM-X.}
    \label{fig:storm-U}
\end{figure}

\begin{figure}[h]
    \centering
    \includegraphics[width=\linewidth]{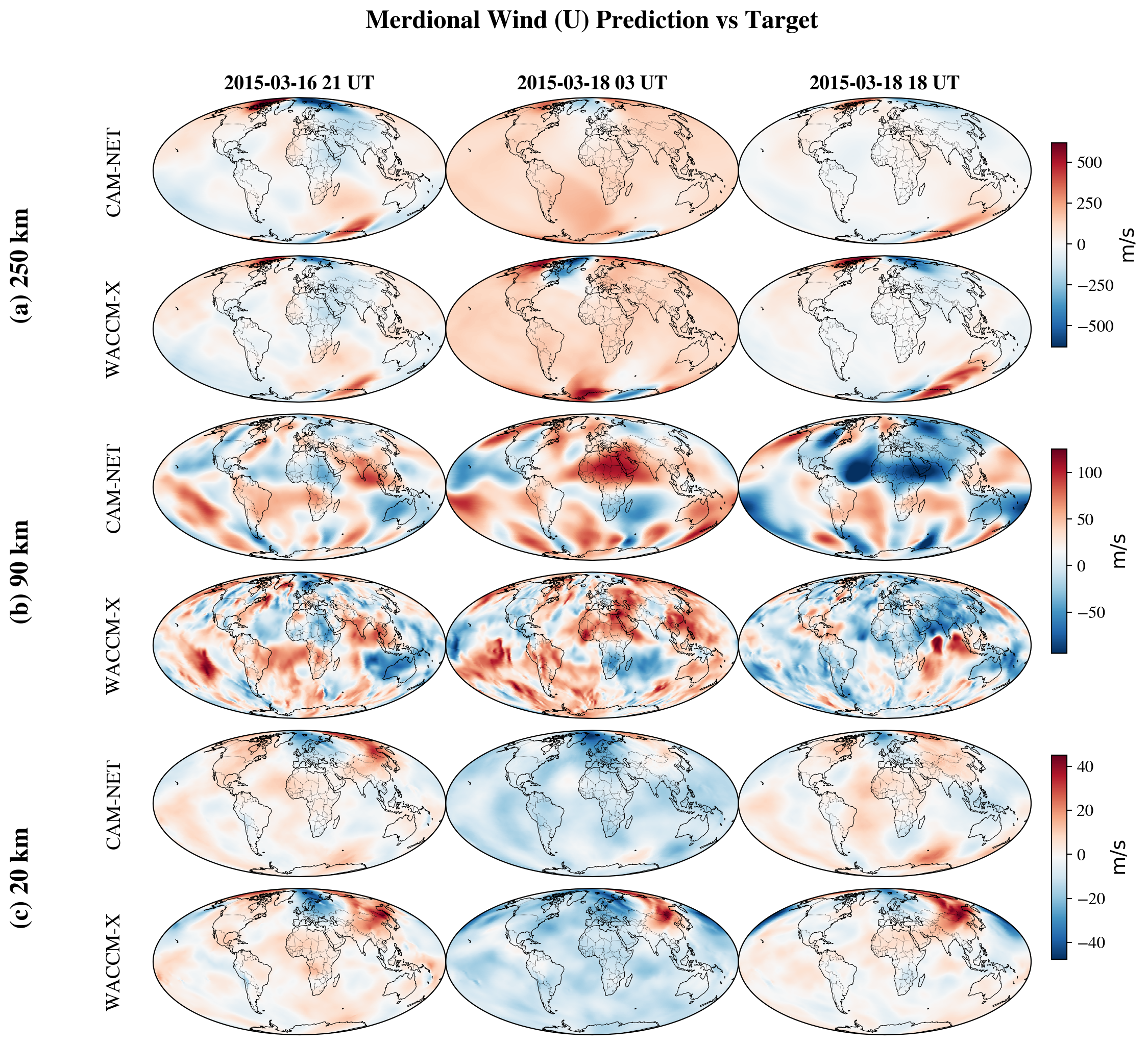}
    \caption{Meridional wind ($V$, m~s$^{-1}$) at approximately (a) 250~km, (b) 90~km, and (c) 20~km. Columns show 21:00~UT on 16 March, 03:00~UT on 18 March, and 18:00~UT on 18 March 2015. Within each panel, the upper row shows CAM-NET and the lower row shows WACCM-X.}
    \label{fig:storm-V}
\end{figure}

Figures~\ref{fig:storm-U} and~\ref{fig:storm-V} compare CAM-NET
predictions with WACCM-X reference fields for zonal wind ($U$) and
meridional wind ($V$), respectively, at approximately 250, 90, and
20~km during the March 2015 geomagnetic storm. Relative to the
quiet-time examples in the main text, the selected storm-time
snapshots exhibit stronger regional asymmetries and more rapidly
varying spatial structure, particularly near 90~km. CAM-NET continues
to reproduce the dominant global-scale circulation, including broad
thermospheric flow patterns at 250~km and coherent lower-atmosphere
structures at 20~km. The largest differences occur near 90~km, where WACCM-X contains
more localized and spatially variable wind structures. CAM-NET
preserves the broad circulation but produces smoother fields with
weaker localized extrema. This behavior is consistent with the
scale-dependent limitation identified in the quiet-time evaluation:
CAM-NET retains planetary-scale and large-scale variability more
successfully than intermittent high-wavenumber mesospheric
structures. Overall, the storm-period wind comparisons indicate that
CAM-NET remains stable at the global scale, while localized
mesospheric variability remains the most challenging regime.



\subsection{Ionospheric variables at 250 km}
\label{app:storm-ionosphere}

\begin{figure}[t]
    \centering
    \includegraphics[width=\linewidth]{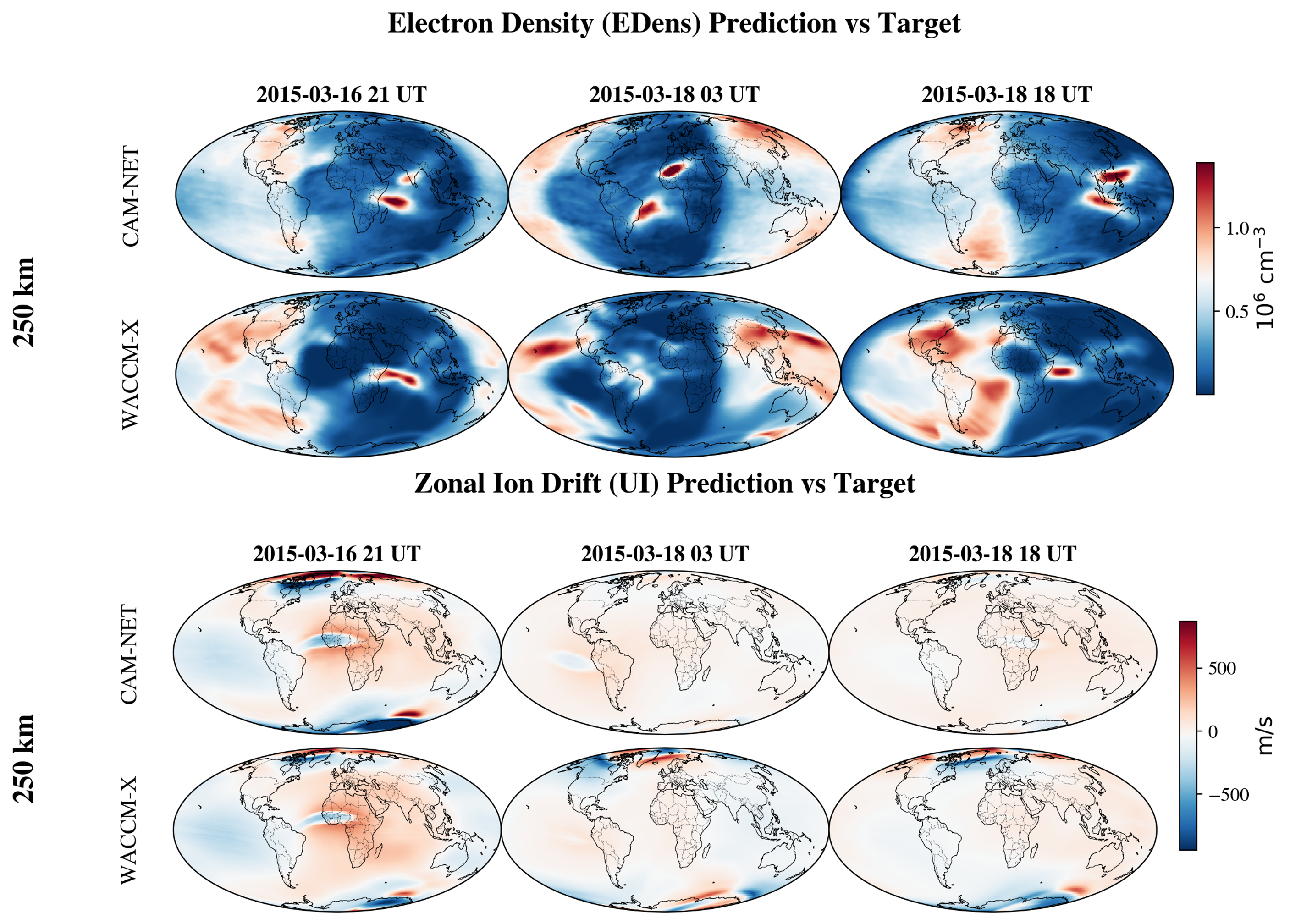}
    \caption{Storm-period comparison of ionospheric electron density ($N_e$; top) and zonal ion drift ($U_i$; bottom) at approximately 250 km during the March 2015 geomagnetic storm. Columns show 21:00 UT on 16 March, 03:00 UT on 18 March, and 18:00 UT on 18 March 2015. For each variable, the upper row shows CAM-NET predictions and the lower row shows WACCM-X reference fields. Electron density is reported in units of $10^{6}\,\mathrm{cm}^{-3}$, and zonal ion drift is reported in $\mathrm{m\,s}^{-1}$.}
    \label{fig:storm-iono}
\end{figure}

Figure~\ref{fig:storm-iono} compares electron density
($N_e$) and zonal ion drift ($U_i$) from CAM-NET and WACCM-X at
approximately 250~km during the March 2015 geomagnetic storm.

For electron density, CAM-NET retains the broad low-latitude and
day--night organization seen in WACCM-X, but the agreement degrades
as the storm evolves. Before the main phase, both models show an
enhancement over the African--Indian Ocean sector, although the
CAM-NET feature is more compact and smoother. During the recovery
phase on 18 March, larger discrepancies emerge in the longitude,
spatial extent, and amplitude of the enhanced-density regions.
CAM-NET places the strongest features mainly over the African--Atlantic
and later the Southeast Asian--western Pacific sectors, whereas
WACCM-X produces broader enhancements extending across the Pacific,
Asian, American, and Indian Ocean sectors. These differences indicate
that CAM-NET preserves part of the large-scale plasma organization but
does not reliably reproduce the timing and longitudinal evolution of
localized storm-time enhancements.

For zonal ion drift, both CAM-NET and WACCM-X show the strongest
variability at high latitudes and broadly similar eastward--westward
polarity patterns. CAM-NET, however, consistently produces smoother
auroral and subauroral structures, weaker extrema, and less distinct
narrow drift bands. As the drift magnitudes decrease on 18 March, the
model retains the broad spatial organization but continues to
attenuate sharp regional gradients. The residual ionospheric module
therefore remains numerically stable during the storm interval, while
showing reduced fidelity for localized electrodynamic structures and
peak amplitudes.

Compared with the quiet-time examples in the main text, the storm-period fields show qualitatively larger differences between CAM-NET and WACCM-X, particularly in the locations and amplitudes of localized extrema. Several factors may contribute to this reduced storm-time fidelity. Although CAM-NET receives Kp and F10.7 as external inputs, these scalar indices provide only bulk measures of geomagnetic and solar activity and do not fully characterize the spatially and temporally varying energy input into the thermosphere-ionosphere system. In WACCM-X, storm-time variability is represented through self-consistent thermosphere-ionosphere electrodynamics, including high-latitude electric-field forcing, ion-neutral coupling, ion drag, Joule heating, and auroral forcing \cite{liu2018development}. These processes depend not only on the overall level of geomagnetic activity but also on the spatial distribution and temporal evolution of magnetospheric forcing. Consequently, periods with similar Kp or F10.7 values can produce substantially different thermosphere-ionosphere responses. Since CAM-NET does not explicitly ingest these spatially distributed external forcing fields, it has limited ability to reconstruct localized storm-time structures from atmospheric states and scalar activity indices alone. Furthermore, strongly disturbed conditions occupy only a relatively small fraction of the training dataset compared with quiet and moderately disturbed periods, further reducing the model's ability to learn rare extreme-event behavior.

\label{app:baseline}

\end{document}